\documentclass[journal]{IEEEtran}
\usepackage[utf8]{inputenc}
\usepackage{bm}
\usepackage{amsmath,epsfig,amssymb,dsfont,amsthm}
\usepackage{hyperref}
\usepackage{verbatim}
\usepackage{caption}
\usepackage{subcaption}
\usepackage{xcolor}
\usepackage{graphicx}
	\graphicspath{{./figures/}}
\usepackage{float}
\usepackage{url}
\usepackage{eurosym}
\usepackage{setspace}
\usepackage{balance}
\usepackage{dsfont}
\usepackage{bbm}
\usepackage{array}
\usepackage{cite}
\usepackage[ruled,vlined]{algorithm2e}
\usepackage{thmtools}
\usepackage{comment}
\usepackage{soul}

\newcommand{\blambda}{\boldsymbol{\Lambda}}
\newcommand{\bA}{\boldsymbol{A}}
\newcommand{\bL}{\boldsymbol{\mathcal L}}
\newcommand{\R}{\mathcal R}
\newcommand{\bT}{\mathsf{T}}
\newcommand{\bE}{\mathbb{E}}

\theoremstyle{plain}
\newtheorem{Thm}{Theorem}

\newtheorem{Prp}{Proposition}
\newtheorem{Lem}{Lemma}
\newtheorem{Asm}{Assumption}
\newtheorem{Pro}{Property}
\theoremstyle{remark}
\newtheorem{Rem}{Remark}

\declaretheoremstyle[%
  spaceabove=-6pt,%
  spacebelow=6pt,%
  headfont=\normalfont\itshape,%
  postheadspace=1em,%
  qed=\qedsymbol%
]{mystyle} 
\declaretheorem[name={Proof},style=mystyle,unnumbered,
]{prf}

\newcommand{\qedsymb}{\hfill\ensuremath{\blacksquare}}                 
\renewcommand{\qedsymbol}{\qedsymb}                                    
\allowdisplaybreaks
\title{Explainability and Graph Learning from Social Interactions}
\author{\IEEEauthorblockN{Valentina Shumovskaia\IEEEauthorrefmark{1}, Konstantinos Ntemos\IEEEauthorrefmark{1}, Stefan Vlaski\IEEEauthorrefmark{2}, and Ali H. Sayed\IEEEauthorrefmark{1}}\\
$\newline$
\IEEEauthorblockA{\IEEEauthorrefmark{1}
École Polytechnique Fédérale de Lausanne (EPFL)
}\\
\IEEEauthorblockA{\IEEEauthorrefmark{2}
Imperial College London}
\thanks{Emails: $\{$valentina.shumovskaia, konstantinos.ntemos, ali.sayed$\}$@epfl.ch, s.vlaski@imperial.ac.uk. This work was supported in part by SNSF grant 205121-184999. A preliminary version of this work without proofs appeared in~\cite{shumovskaia}.}
}

\IEEEoverridecommandlockouts
\begin{document}

\maketitle

\begin{abstract}
    Social learning algorithms provide models for the formation of opinions over social networks resulting from local reasoning and peer-to-peer exchanges. Interactions occur over an underlying graph topology, which describes the flow of information among the agents. To account for drifting conditions in the environment, this work adopts an adaptive social learning strategy, which is able to track variations in the underlying signal statistics. Among other results, we propose a technique that addresses questions of explainability and interpretability of the results when the graph is hidden. Given observations of the evolution of the beliefs over time, we aim to infer the underlying graph topology, discover pairwise influences between the agents, and identify significant trajectories in the network. The proposed framework is online in nature and can adapt dynamically to changes in the graph topology or the true hypothesis.
\end{abstract}


\section{Introduction}
    This work focuses on the \textit{social learning} paradigm~\cite{jadbabaie2012non, nedic2017fast, molavi2017foundations, molavi2018theory, bordignon2020adaptive, bordignon2020social, lalitha2018social, zhao2012learning, gale2003bayesian, acemoglu2011bayesian}, where agents react to streaming data and information shared with their neighbors. Social learning refers to the problem of distributed hypothesis testing, where each agent aims at learning an underlying true hypothesis (or state) through its own observations and from information shared with its neighbors.
    Social learning studies can be categorized into Bayesian \cite{gale2003bayesian, acemoglu2011bayesian} and non-Bayesian \cite{jadbabaie2012non, nedic2017fast, molavi2017foundations, molavi2018theory, bordignon2020adaptive, bordignon2020social, lalitha2018social, zhao2012learning}. Non-Bayesian approaches have gained increased interest due to their appealing scalability traits. In these approaches, agents follow a two-stage process at every time instant. First, every agent updates its belief (a probability distribution over the possible hypotheses) based on its currently received private observation from the environment. Then, it fuses the shared beliefs from its neighbors. The main focus of these studies is to prove that agents' beliefs across the network converge to the true hypothesis after sufficient repeated interactions. We are interested in studying a dynamic setting where both the graph topology and the true hypothesis can change over time. For these reasons, we adopt the general adaptive social learning protocol from~\cite{bordignon2020adaptive}.
    
    In this work, we aim at revealing the underlying influence pattern over a social learning network. By doing so, we discover the influences that drive opinion formation in the network and identify the most significant trajectories for information propagation as well as the most relevant nodes. We propose techniques that enable the designer to address useful questions related to explainability and interpretability~\cite{baehrens2010explain, adadi2018peeking, gilpin2018explaining, BARREDOARRIETA202082} over graphs. 
    
    The explainability problem is relevant for multi-agent networks and it has been receiving increased attention~\cite{heuillet2022collective, ohana2021explainable, 9355456}. While cooperation is beneficial over such networks, leading to more stable performance, it nevertheless makes interpretability of the results more challenging especially when the underlying topology is unknown to the observer. It therefore becomes critical to learn which agents contribute to learning the most and to estimate how agents influence each other. Motivated by these considerations, in this work we examine the problem of explainability over graphs driven by social interactions. While we focus on social learning as a model for social interactions in this work, some of the ideas and results can be extended to other settings, such as estimation, optimization, or multi-agent reinforcement learning~\cite{vlaski22, wang2019influence, Sayed_2014, Sayed_2023}.
    
    While the effect of observations in (non-)Bayesian inference is fairly well-understood, the effect of the network on decision making is less understood. Even more, the topology itself is generally unknown. Full understanding and, hence, interpretability of the social network's decision requires learning the graph topology as a key step. For this purpose, we will devise an online algorithm for recovering the weights over the edges that determine the graph. In the context of social learning, recovering the graph helps provide important insight into the decision-making process. 
    
    A number of solutions for graph learning~\cite{kalofolias2016learn, egilmez2017graph, vlaski2018online, dong2019learning, pasdeloup2017characterization, thanou2017learning, chepuri2017learning, shafipour2017network, segarra2016network, viola2018graph, sardellitti2016graph, maretic2017graph, shi2019bayesian, shahrampour2013reconstruction, hassan2016topology, liu2019dynamic} have already been proposed in the literature, where algorithms for graph inference have been developed for particular models, describing the relationship between observations and graphs. For instance, graph learning for the heat diffusion process is studied in~\cite{pasdeloup2017characterization, vlaski2018online, thanou2017learning, ma2008mining}, while learning under structural constraints, such as connectivity~\cite{egilmez2017graph} and sparsity, appears in~\cite{egilmez2017graph, chepuri2017learning, maretic2017graph}, and approaches based on examining the precision matrix appear in  \cite{friedman2008sparse, matta2019graph}. Most of these works consider \textit{static} graphical models. This is in contrast to graphs with dynamic properties~\cite{vlaski2018online} where the connectivity among agents can change over time. The approach pursued in this work is applicable to dynamic scenarios. Moreover, and importantly, it does not require knowledge of the true state. Once the graph is learned, we will then devise a procedure for identifying the most influential agents and trajectories. 
    
    The manuscript is organized as follows. We describe the system model in Section~\ref{sec:system_model}, and derive the online graph learning algorithm in Section~\ref{sec:algorithm}. In Section~\ref{sec:explain}, we discuss explainability and interpretability and propose a technique to quantify pairwise influences and trajectories. In Section~\ref{sec:experiments}, we provide experiments to illustrate the robustness of the graph learning method against dynamic changes to the topology and the true hypothesis, and to reveal how influences are identified. 

\section{Social Learning Model}
\label{sec:system_model}
    We consider a set $\mathcal{N}$ of agents connected by a directed graph $\mathcal{G}=\left( \mathcal{N}, \mathcal{E} \right)$, where $\mathcal{E}$ represents the links between agents. 
    An agent $k \in \mathcal N$ can share the information directly with another agent $\ell \in \mathcal N$ when the link $(\ell, k)$ is present in the set $\mathcal E$.
    The set of neighbors of an agent $k\in\mathcal{N}$ including itself, is denoted by $\mathcal{N}_k$.
    
    All agents aim at learning the true hypothesis $\theta^{\star}$, belonging to a finite set of possible hypotheses denoted by $\Theta$ (whose cardinality is at least two). To this end, each agent $k$ receives streaming observations $\boldsymbol{\zeta}_{k,i}\in\mathcal{Z}_k$ at every time $i\geq1$, where $\mathcal{Z}_k$ is a compact set. Agent $k$ also has access to the likelihood functions $L_k(\boldsymbol{\zeta}_{k,i}|\theta)$, for all $\theta\in\Theta$. The signals $\boldsymbol{\zeta}_{k,i}$ are independent over both time and space, and are also identically distributed (i.i.d.) over time. We will use the notation $L_k(\theta)$ instead of $L_k(\boldsymbol{\zeta}_{k,i}\vert\theta)$ for brevity. At each time $i$, agent $k$ keeps a {\em belief vector} $\boldsymbol{\mu}_{k,i}$, which is a probability distribution over the possible states. The belief component $\boldsymbol{\mu}_{k,i}(\theta)$ quantifies the confidence by agent $k$ that $\theta$ is the true state. Therefore, at time $i$, each agent's true state estimator is computed as follows:
    \begin{align}
        \widehat{\boldsymbol{\theta}}_{k,i}^\circ = \arg\max_{\theta\in\Theta}\boldsymbol{\mu}_{k,i}(\theta).
        \label{eq:truestate_est0}
    \end{align}  
    To avoid technicalities and to ensure well-posedness and identifiability, we need to impose certain common conditions on the graph topology and on the initial beliefs. For instance, the following statement ensures that agents would not discard any particular state a-priori.
    \begin{Asm}[\bf{Positive initial beliefs}]
        \label{positive_beliefs}
            For all hypotheses $\theta\in\Theta$, all agents $k\in\mathcal N$ start with positive initial beliefs, $\boldsymbol{\mu}_{k,0}(\theta)>0$.
        \qedsymb
        \label{asm:beliefs}
    \end{Asm}
    At every time instant $i$, every agent $k$ updates its belief by using a two-stage process. First, it incorporates information from the received observation $\boldsymbol{\zeta}_{k,i}$ and then it fuses the information from its neighbors. More specificially, in this work we consider the \textit{adaptive social learning} rule~\cite{bordignon2020adaptive}, which has been shown to have favorable transient and steady-state performance in terms of convergence rate and probability of error for dynamic environments. Under this protocol, agents update their beliefs in the following manner:
    \begin{align}
        &\boldsymbol{\psi}_{k,i}(\theta)=
        \frac{L_k^\delta(\boldsymbol{\zeta}_{k,i}|\theta)\boldsymbol{\mu}^{1-\delta}_{k,i-1}(\theta)}{\sum_{\theta'\in\Theta}L_k^{\delta}(\boldsymbol{\zeta}_{k,i}|\theta')\boldsymbol{\mu}^{1-\delta}_{k,i-1}(\theta')},\quad k\in\mathcal{N}\label{eq:adapt_adaptive}\\
        \nonumber\\
        &\boldsymbol{\mu}_{k,i}(\theta)=\frac{\prod_{\ell\in\mathcal{N}_k}\boldsymbol{\psi}^{a_{\ell k}}_{\ell,i}(\theta)}{\sum_{\theta'\in\Theta}\prod_{\ell\in\mathcal{N}_k}\boldsymbol{\psi}^{a_{\ell k}}_{\ell,i}(\theta')}, \quad k\in\mathcal{N} \label{eq:combine}
    \end{align}
    where $a_{\ell k}$ denotes the {\em combination weight} assigned by agent $k$ to neighboring agent $\ell$, satisfying $0<a_{\ell k}\leq1$, for all $\ell\in\mathcal{N}_k$, $a_{\ell k}=0$ for all $\ell\notin\mathcal{N}_k$, and $\sum_{\ell\in\mathcal{N}_k}a_{\ell k}=1$. The algorithm is called ``\textit{adaptive}'' due to the parameter $\delta \in (0,1)$, which allows it to track changes in the true hypothesis $\theta^\star$. Observe that the numerator in~(\ref{eq:combine}) is the weighted geometric mean of the priors $\bm{\psi}_{\ell,i}(\theta)$ at time $i$ with weights given by the scalars $\{a_{\ell k}\}$. We further assume that information is able to flow throughout the network by requiring strong-connectedness in the manner defined next~\cite{Sayed_2014, Sayed_2023}.
    \begin{Asm}[\bf{Strongly-connected network}]
        \label{strongly_connected}
            The communication graph is {\em strongly connected}. That is, there exists a path with positive weights linking any two agents, and at least one agent in the graph has a self-loop, meaning that there is at least one agent $k\in\mathcal{N}$ with $a_{kk}>0$.
        \qedsymb
        \label{asm:network}
    \end{Asm}
    
    The agents run recursions (\ref{eq:adapt_adaptive})--(\ref{eq:combine}) continually. However, the observer can only track the shared beliefs $\{\bm{\psi}_{k,i}(\theta)\}$. The graph and its combination weights are not known. Thus, let $A_{\star}=[a_{\ell,k}]$  denote the true (yet unknown) left-stochastic {\em combination matrix} consisting of all combination weights $a_{\ell k}$. It is known that the power matrix  $(A_\star^{\mathsf{T}})^t$ converges to $\mathds{1}u^{\mathsf{T}}$ at an exponential rate governed by the second largest-magnitude eigenvalue of $A_\star$, as $t\rightarrow\infty$~\cite{sayed2014diffusion}, where $u$ denotes the Perron eigenvector, i.e., $A_{\star}u=u$, with entries satisfying $\sum_\ell u_{\ell} = 1$ and  $u_\ell > 0$ for $\ell \in \mathcal N$. The following property holds for such matrices~\cite{horn2009}.
        \begin{Pro}[\bf{Closeness to Perron entries}]
            Let $\beta_2$ be the second largest-magnitude eigenvalue of $A_\star$. Then, for any positive $\beta$ such that $|\beta_2| < \beta < 1$, there exists a positive constant $\sigma$ (depending only on $A_\star$ and $\beta$), such that, for all $\ell, k \in \mathcal N$, and for all $t = 1,2,\dots$, we have that:
            \begin{align}
                \big| [A^t_\star]_{\ell, k} - u_\ell \big| \leq \sigma \beta^t.
            \end{align}
            \label{property}
            \qedsymb
        \end{Pro}
    \noindent In addition, for well-posedness, we will assume that the agents can collectively identify the underlying true hypothesis~\cite{lalitha2016social, bordignon2020adaptive}. In other words, agents are able to distinguish collectively any hypothesis $\theta\in\Theta$ from $\theta^\star$. This requirement amounts to the following identifiability condition. 
    \begin{Asm}[\bf{Identifiability assumption}]
            For each wrong hypothesis $\theta\neq\theta^\star$, there is at least one agent $k\in \mathcal N$ that has strictly positive KL-divergence $D_{KL}\left(L_k\left(\theta\right)||L_k\left(\theta^\star\right)\right) > 0$.
        \newline $\textrm{ }$\qedsymb
        \label{asm:ident}
    \end{Asm}
    \noindent We also discard degenerate situations by imposing a boundedness condition on the likelihood functions, in a manner similar to~\cite{bordignon2020social}.
    \begin{Asm}[\bf{Bounded likelihoods}]
        \label{asm:support}
        There exists a finite constant $b > 0$ such that, for all $k \in \mathcal N$:
        \begin{align}
            \Bigg|\log \frac {L_k(\boldsymbol\zeta | \theta)}{L_k(\boldsymbol\zeta | \theta')} \Bigg| \leq b
        \end{align}
        for all $\theta,\;\theta' \in \Theta$, and $\boldsymbol\zeta \in \mathcal Z_k$.
        \qedsymb
    \end{Asm}
    \noindent This assumption implicitly assumes that the likelihood models $L_k(\theta)$ have a common support for different hypotheses $\theta\in\Theta$, and that their positive values are bounded away from zero~\cite{nedic2017fast}.
    
\section{Inverse Modeling Problem}
\label{sec:algorithm}
    \subsection{Problem Statement}
        In our study, we assume that the graph is completely hidden. The assumption is motivated by the fact that in real-world settings, the pattern of interactions among agents is usually unknown to an external observer. In addition, in the social learning paradigm, it is common~\cite{lalitha2016social, lalitha2018social} to assume that the local observations $\boldsymbol{\zeta}_{k,i}$ are private and external observers do not have access to them. On the other hand, beliefs (i.e., $\boldsymbol{\psi}_{k,i}(\theta)$) are public and exchanged over edges across the network. For this reason, our goal is to infer the graph topology by observing the exchanged beliefs among the agents.
        
        Formally, we assume that at each time step $i\geq 1$ we observe the beliefs of the agents in the network, collected into the set:
        \begin{align}
        \label{eq:obs_information}
            {\mathcal D}_i = \Bigl\{ \bm{\psi}_{k,i}(\theta),\;k\in{\mathcal N}\Bigr\}
        \end{align}
        The problem of interest is to recover the combination matrix $A_\star$ based on knowledge of $\{\mathcal D_i\}_{i\geq 1}$.
        
    \subsection{Likelihood and Beliefs Ratios}
        We introduce two matrices $\blambda_i$ and $\bL_i$ of size $|\mathcal N| \times (|\Theta|-1)$, where each element in these matrices is a relative measure of log beliefs and likelihood ratios as follows:
        \begin{align}
            &[\boldsymbol{\Lambda}_{i}]_{k,j} \triangleq\log\frac{\boldsymbol{\psi}_{k,i}(\theta_0)}{\boldsymbol{\psi}_{k,i}(\theta_j)}
            \label{eq:lambda}\\
            &[\boldsymbol{\mathcal{L}}_{i}]_{k,j}\triangleq\log\frac{L_k(\boldsymbol{\zeta}_{k,i}\vert\theta_0)}{L_k(\boldsymbol{\zeta}_{k,i}\vert\theta_j)}.
            \label{eq:loglikelihood}
        \end{align}
        In these expressions, we have chosen some arbitrary $\theta_0\in\Theta$ as a reference state, while $\theta_j\neq \theta_0$. 
        
        Observe that both matrices in (\ref{eq:lambda})--(\ref{eq:loglikelihood}) vary with  the time index $i$. Based on the definitions~(\ref{eq:lambda})--(\ref{eq:loglikelihood}), some algebra~(see Appendix~\ref{app:recursion}) will show that we can transform~(\ref{eq:adapt_adaptive})--(\ref{eq:combine}) into an update relating these matrices:
        \begin{align}
            &\boldsymbol{\Lambda}_i=(1-\delta)A_\star^{\mathsf{T}}\boldsymbol{\Lambda}_{i-1}+\delta\boldsymbol{\mathcal{L}}_i.
            \label{eq:recursion_adaptive}
        \end{align}
        
        \noindent Due to Assumption~\ref{asm:support}, $\bL_i$ and $\blambda_i$ have bounded entries at each iteration $i$ and in the limit as $i\rightarrow\infty$.
        
        \begin{Lem}[\bf{Asymptotic properties of log-belief matrix}]
            Under Assumptions~\ref{asm:beliefs}-\ref{asm:support},
            the random matrices $\blambda_i$ converge to the following random variable $\blambda$ in distribution as $i\rightarrow \infty$:
            \begin{align}
                \blambda \triangleq \delta \sum_{i=0}^\infty \left(1-\delta\right)^{i}(A_\star^{i})^\bT\bL_i.
            \end{align}
            Moreover, for every $i \geq 1$, $\blambda_i$ is elementwise bounded (in absolute value) by a matrix $\bar{\Lambda}_i$, written as follows:
            \begin{align}
                |\blambda_i| \preceq \bar\Lambda_i
            \end{align}
            where
            \begin{align}
                \bar\Lambda_i \triangleq \left(1-\delta\right)^i \left(A_\star^\bT\right)^i|\Lambda_0| + \delta b \sum_{t=0}^{i-1} \left(1-\delta\right)^{t}(A_\star^{t})^\bT \mathds{1}\mathds{1}^\bT.
            \end{align}
            In the limit,
            \begin{align}
                \bar \Lambda \triangleq \lim_{i\rightarrow\infty} \bar \Lambda_i = \delta b \sum_{t=0}^\infty \left(1-\delta\right)^{t}(A_\star^{t})^\bT \mathds{1}\mathds{1}^\bT.
            \end{align}
            \label{lem:lambda}
        \end{Lem}
        \begin{prf}
            See Appendix \ref{l1}.
        \end{prf}
        
        At every iteration $i$, the quantities $\{\blambda_i,\blambda_{i-1}\}$ are known based on knowledge of the beliefs ${\mathcal D}_i$ from (\ref{eq:obs_information}). On the other hand, the quantity $\bL_i$ is not known because the observations $\{\bm{\zeta}_{k,i}\}$ are private.  We wish to devise a scheme that allows us to estimate $A_{\star}$ in (\ref{eq:recursion_adaptive}) from knowledge of $\{\blambda_i,\blambda_{i-1}\}$ and from a suitable approximation for $\bL_i$. Before discussing the learning algorithm, however, we establish the following useful property. For simplicity of notation, we will write 
        \begin{align}
            \mathbb E [\cdot] \triangleq \mathbb E_{\boldsymbol\zeta_{k,t}\sim L_k(\theta^\star), k\in\mathcal N, t\leq i} [\cdot]
        \end{align}
        where the expectation is relative to the randomness in all local observations up to time $i$. 
        
        \begin{Prp}[\bf{Mean likelihood matrix}]
            The random matrices $\bL_i$ are i.i.d. over time and space, and their mean matrix $\bar\bL=\bE\bL_i$ is independent of time and finite with individual entries:
            \begin{align}
                [\bar \bL]_{k,j} =\textrm{ }&  D_{KL}\left(L_k\left(\theta^\star\right)||L_k\left(\theta_j\right)\right) \nonumber\\
                &- D_{KL}\left(L_k\left(\theta^\star\right)||L_k\left(\theta_0\right)\right).
                \label{eq:L_exp}
            \end{align}
            \label{lem:loglikelihoods}
        \end{Prp}
        \begin{prf}
            Since the private data $\boldsymbol\zeta_{k,i}$ is i.i.d. and follows distribution $L_k(\theta^\star)$, the expectation of $\bL_i$ does not depend on $i$ and is equal to:
            \begin{align}
                [\bar\bL]_{k,j} 
                =\textrm{ }& \bE \log L_k(\boldsymbol\zeta_{k,i}|\theta_0) - \bE \log L_k(\boldsymbol\zeta|\theta_j) \nonumber\\
                =\textrm{ }&  D_{KL}\left(L_k\left(\theta^\star\right)||L_k\left(\theta_j\right)\right) \nonumber\\
                &- D_{KL}\left(L_k\left(\theta^\star\right)||L_k\left(\theta_0\right)\right).
            \end{align}
            Under Assumption~\ref{asm:support}, the entries of $\bar\bL$ are bounded.
            $\newline$
        \end{prf}
        
    \subsection{Algorithm Development}
        To motivate the algorithm, we assume first for the sake of the argument that the observer knows the true state $\theta^\star$, and we remove this assumption further ahead.

        The linear nature of the update for $\blambda_i$ in~(\ref{eq:recursion_adaptive}) motivates the following instantaneous quadratic loss function for finding $A_\star$:
        \begin{align}
            Q'(A; \blambda_i, \blambda_{i-1}) = \frac{1}{2} \| \boldsymbol{\Lambda}_i -(1-\delta)A^{\mathsf{T}}\boldsymbol{\Lambda}_{i-1} - \delta \boldsymbol{\mathcal{L}}_i \|_{\rm F}^2,
            \label{eq:cost0}
        \end{align}
        where $\|\cdot\|_{\rm F}$ is the Frobenius norm. Computation of $\bL_i$ requires knowledge of $\boldsymbol\zeta_{k, i}$, $k\in\mathcal N$, which is private for each agent and is therefore hidden from the observer. For this reason, we will assume instead knowledge of $\bar \bL$, which still requires knowledge of the true hypothesis $\theta^\star$. We explain in the sequel how to circumvent this latter requirement. Using $\bar\bL$, we replace~(\ref{eq:cost0}) by
        \begin{align}\label{eq:cost_function_adaptive} 
            &Q(A; \blambda_i, \blambda_{i-1}, \bar \bL) = \frac{1}{2} \| \boldsymbol{\Lambda}_i -(1-\delta)A^{\mathsf{T}}\boldsymbol{\Lambda}_{i-1} - \delta \bar \bL \|_{\rm F}^2.
        \end{align}
        and use this loss function to introduce an optimization problem over a horizon of $N$ time instants for the recovery of $A_{\star}$ as follows:
        \begin{align}
        \label{eq:min_problem}
            &\min_{A} J(A) \triangleq \frac 1N\sum_{i=1}^N J_i(A),\\
            &J_i(A) \triangleq \bE Q(A;\blambda_{i}, \blambda_{i-1}, \bar\bL).
         \label{eq:risk}
        \end{align}
        The statistical properties of $\blambda_i$ vary with time, and this explains why we are averaging over a time-horizon in~(\ref{eq:min_problem}).
        We will apply stochastic approximation to solve~(\ref{eq:min_problem}) and use a recursion of the form:
        \allowdisplaybreaks[0]
        \begin{align}
            \boldsymbol{A}^{\mathsf{T}}_i =&\textrm{ } \boldsymbol{A}^{\mathsf{T}}_{i-1}
            + \mu(1-\delta) \nonumber\\
            \;\times&\left(\boldsymbol{\Lambda}_i -  (1-\delta)\boldsymbol{A}_{i-1}^{\mathsf{T}}\boldsymbol{\Lambda}_{i-1} - \delta \bar \bL \right) \boldsymbol{\Lambda}_{i-1}^{\mathsf{T}}
            \label{eq:graph_update_unbiased}
        \end{align}
        \allowdisplaybreaks
        
        \begin{Lem}[\bf{Risk function properties}]
            Each risk function $J_i(A)$ defined by~(\ref{eq:risk}) is strongly convex and has Lipschitz gradient with corresponding constants $\nu_i$ and $\kappa_i$  defined in terms of the smallest and largest eigenvalues of the second-order moment matrix $\bE\blambda_{i-1}\blambda_{i-1}^\bT$:
            \begin{align}
                &\nu_i = (1-\delta)^2\lambda_{\min}\left(\bE\blambda_{i-1}\blambda_{i-1}^\bT\right)
                \label{eq:nu}\\
                &\kappa_i = (1-\delta)^2\lambda_{\max}\left(\bE\blambda_{i-1}\blambda_{i-1}^\bT\right)
                \label{eq:kappa}
            \end{align}
            Also, $A_\star$ is the unique minimizer for $J_i(A)$.
            \label{lem:cost}
        \end{Lem}
        \begin{prf}
            See Appendix \ref{l2}.
        \end{prf}
        \noindent
        In order to examine the steady-state performance of recursion~(\ref{eq:graph_update_unbiased}), we introduce an independence assumption that is common in the study of adaptive systems~\cite{Sayed_2014}.
        \begin{Asm}[\bf{Separation principle}]
            \label{asm:independence}
            Let $\widetilde{\bA}_i=A_\star-\bA_i$ denote the estimation error. Assume the step-size $\mu$ is sufficiently small so that  $\|\widetilde{\bA}_{i}\|_{\rm F}^2$ reaches a steady state distribution and $\widetilde{\bA}_i$ is independent of the observation $\blambda_i$ in the limit, conditioned on the history of past observations prior to time $i$.
            \qedsymb
        \end{Asm}
        \noindent We also assume positive-definite second-order moments for the matrices $\bL_i$, which are i.i.d. over time.
        \begin{Asm}[\bf{Positive-definite second moments}]
            \label{asm:loglikelihoods_positivedefinite}
            At any moment $i \geq 0$ and for any chosen $\theta_0 \in \Theta$, the second order moment of matrix $\bL_i$ in~(\ref{eq:loglikelihood}) is uniformly positive-definite, $\bE\bL_i\bL_i^\bT \succeq \tau I$ for some $\tau > 0$ for any $i$.
            \qedsymb
        \end{Asm}
        \noindent Since the observations $\boldsymbol{\zeta}_{k,i}$ are independent among agents, and are also identically distributed (i.i.d.) over time, the above assumption holds if the variances for each individual agent $k\in\mathcal N$ satisfies:
        \begin{align}
            \sum_j \mbox{\rm Var}\left([\boldsymbol{\mathcal L}_i]_{k,j}\right) = \sum_j \mbox{\rm Var}\left(\log \frac {L_k(\boldsymbol\zeta_{i,k}|\theta_j)}{L_k(\boldsymbol\zeta_{i,k}|\theta_0)}\right) \geq \tau > 0.
        \end{align}
        
        Using these conditions, we can establish the following steady-state result, which shows that the mean-square error approaches $O(\mu)$.
        \begin{Thm}[\bf{Steady-state performance}]
            Under Assumptions
            \ref{asm:beliefs}-\ref{asm:loglikelihoods_positivedefinite}, the mean-square deviation (MSD) converges exponentially fast with
            \begin{align}
                \limsup_{i\rightarrow\infty} \bE \|\widetilde{\bA}_{i}\|_{\rm F}^2
                \leq \frac{\mu^2\gamma}{1-\alpha} = O(\mu),
            \end{align}
            where
            \begin{align}
                \alpha &= 1 - 2\mu\nu + O(\mu^2) \nonumber\\
                \gamma &=\delta^2\kappa |\mathcal N|\lambda_{\max}(\R_{\bL})\nonumber\\
                \nu &= (1-\delta)^2\lambda_{\min}\left(\lim_{i\rightarrow\infty}\bE\blambda_{i}\blambda_{i}^\bT\right)\nonumber\\
                \kappa &= (1-\delta)^2\lambda_{\max}\left(\lim_{i\rightarrow\infty}\bE\blambda_i\blambda_{i}^\bT\right)
            \end{align}
            and $\R_{\bL} \triangleq \bE (\bL_i - \bar \bL)(\bL_i - \bar\bL)^\bT$ is independent of $i$.
            \label{thm:conv_0}
        \end{Thm}
        \begin{prf}
            See Appendix \ref{t1}.
        \end{prf}
    
    \subsection{True State Learning}
        \label{sec:true_state_learning}
        In the previous section, we assumed knowledge of the true state $\theta^\star$, which is needed to evaluate $\bar\bL$. We now relax this condition and generalize the algorithm.
        
        Typically, at each time step $i\geq 1$, every agent $k\in\mathcal N$ estimates the true state using~(\ref{eq:truestate_est0}).
        It was shown in \cite[Theorem 2]{bordignon2020adaptive}) that the probability of error tends to zero, i.e., $\mathbb P (\widehat{\boldsymbol{\theta}}_{k,i}^\circ \neq \theta^\star) \rightarrow 0$ as $i\rightarrow\infty$ and $\delta \rightarrow 0$. The same conclusion continues to hold if we replace (\ref{eq:truestate_est0}) and estimate the underlying hypothesis based on the intermediate belief vectors (which are the quantities that are assumed to be observable):
        \begin{align}
            \widehat{\boldsymbol{\theta}}_{k,i} = \arg\max_{\theta\in\Theta}\boldsymbol{\psi}_{k,i}(\theta).
        \end{align}
        The result is summarized in Lemma~\ref{lem:belief}.
        \begin{Lem}[\bf{True state learning error}]
            The error probability for all agents $k \in \mathcal N$ converges to zero as $i\rightarrow\infty$ and step-size $\delta\rightarrow 0$:
            \begin{align}
                \lim_{\delta \rightarrow 0} \mathbb P (\lim_{i\rightarrow\infty}\widehat{\boldsymbol\theta}_{k,i} \neq \theta^\star) = 0.
            \end{align}
            \label{lem:belief}
        \end{Lem}
        \begin{prf}
            See Appendix \ref{l3}.
        \end{prf}
        \noindent In order to agree on a single estimate for  $\theta^{\star}$ among the agents, we will estimate a common $\widehat{\boldsymbol{\theta}}_{i} $ by using a majority vote rule. Then, the following conclusion holds.
        \begin{Lem}[\bf{True state learning error: majority vote}]
            \begin{align}
                \lim_{\delta\rightarrow0}\mathbb P \left(\lim_{i\rightarrow\infty}\widehat {\boldsymbol\theta}_{i} \neq \theta^\star\right)  = 0.
            \end{align}
            \label{lem:majvote}
        \end{Lem}
        \begin{prf}
            Consider
            \begin{align}
                &\mathbb P \left(\widehat{\boldsymbol\theta}_i \neq \theta^\star\right)  \nonumber\\
                \leq &\;
                \sum_{n = \lceil \frac {|\mathcal N|}2 \rceil}^{|\mathcal N|} \mathbb P \left( \exists k_1,\dots,k_n \in \mathcal N \colon \widehat{\theta}_{k_1,i} \neq \theta^\star,\dots, \widehat{\theta}_{k_n,i} \neq \theta^\star \right) \nonumber\\
                =&\; \sum_{n = \lceil \frac {|\mathcal N|}2 \rceil}^{|\mathcal N|} \left( \frac{|\mathcal N|}{k_1\dots k_n}\right) \mathbb P \left(  \widehat{\theta}_{k_1,i} \neq \theta^\star\right) \dots \mathbb P \left( \widehat{\theta}_{k_n,i} \neq \theta^\star\right)
            \end{align}
            Using the result of Lemma~\ref{lem:belief}, we finish the proof.\\
        \end{prf}
        
        We now introduce a revised optimization problem, where the loss function is based on the estimated hypothesis through the majority vote:
        \begin{align}
        \label{eq:min_problem2}
            \min_{A} \widehat J(A) \triangleq \frac 1N\sum_{i=1}^N \bE Q(A;\blambda_{i}, \blambda_{i-1}, \bar\bL_i)
        \end{align}
        where the original $\bar{\bL}$ is replaced by $\bar\bL_i = \bE_{\widehat{\boldsymbol\theta}_i}\bL_i$ with the expectation now computed relative to the distribution of $\widehat{\boldsymbol\theta}_i:$ $\boldsymbol{\zeta}_{k,i} \sim L_k(\boldsymbol{\zeta}_{k,i}|\widehat{\boldsymbol\theta}_i)$.
        
        We summarize the recursions of the proposed method in Algorithm~\ref{alg}, referred to as the Online Graph Learning (OGL) algorithm. The listing includes three steps: true state estimation for each agent, majority vote on the true state, and the graph update. We obtain the following steady-state performance for the algorithm.
        
        \begin{algorithm}
            \KwData{
                At each time $i,\dots,N$
                \begin{align*}
                    \{ \mathcal D_i,\;\delta,\; D_{KL}(L_k(\theta)||L_k(\theta')),\;k \in \mathcal N,\; \theta,\;\theta' \in \Theta
                    \}
                \end{align*}
            }
            \KwResult{Estimated combination matrix $\bA_N$}
            initialize $\bA_0$\\
            $i = 1$\\
            \Repeat{sufficient convergence}{
                \For{$k\in\mathcal N$}{
                    Every agent estimates the true state individually:
                    \begin{flalign*}
                        &\widehat {\boldsymbol{\theta}}_{k,i} = \arg\max_{\theta\in\Theta}\boldsymbol\psi_{k,i}(\theta)&&
                    \end{flalign*}
                    }
                A majority rule is applied across agents:
                \begin{flalign*}
                    &\widehat {\boldsymbol\theta}_i = \arg\max_\theta \sum_{k \in \mathcal N} \mathds 1\left\{\widehat {\boldsymbol\theta}_{k,i} = \theta\right\}&&
                \end{flalign*}
                
                \noindent Compute matrices $\blambda_i$, $\bar\bL_i$:\\
                \For{$k\in\mathcal N$, $j=1,\dots,|\Theta|$}{
                    \begin{flalign*}
                        &[\boldsymbol{\Lambda}_{i}]_{k,j} = \log\frac{\boldsymbol{\psi}_{k,i}(\theta_0)}{\boldsymbol{\psi}_{k,i}(\theta_j)}&&\\
                        &[\bar\bL_i]_{k,j} = D_{KL}\left(L_k(\widehat {\boldsymbol\theta}_i)||L_k\left(\theta_j\right)\right)&&\\
                        &\textrm{ }\textrm{ }\textrm{ }\textrm{ }\textrm{ }\textrm{ }\textrm{ }\textrm{ }\textrm{ }\textrm{ }\textrm{ } - D_{KL}\left(L_k(\widehat {\boldsymbol\theta}_i)||L_k\left(\theta_0\right)\right)&&
                    \end{flalign*}
                }
                Combination matrix update:
                \begin{flalign*}
                    &\boldsymbol{A}_i =  \boldsymbol{A}_{i-1} + \mu(1-\delta)\blambda_{i-1} \nonumber&&\\
                    \times&\left(\blambda_i^{\mathsf{T}} -  (1-\delta)\blambda_{i-1}^{\mathsf{T}}\boldsymbol{A}_{i-1} - \delta \bar\bL_i\right).&&
                \end{flalign*}
                \begin{flalign*}
                    &i = i+1&&
                \end{flalign*}
            }
            \caption{Online Graph Learning (OGL)}
            \label{alg}
        \end{algorithm}
        
        \begin{Thm}[\bf{Steady-state performance}]
            Under Assumptions \ref{asm:beliefs}-\ref{asm:loglikelihoods_positivedefinite}, after large enough number of social learning iterations with $\delta \rightarrow 0$ and for sufficiently small $\mu$, the mean-square deviation (MSD) converges exponentially fast with:
            \begin{align}
                \limsup_{i\rightarrow\infty} \bE \|\widetilde{\bA}_{i}\|_{\rm F}^2
                \leq \frac{\mu^2\gamma}{1-\alpha} = O(\mu).
            \end{align}
            \label{thm:conv}
        \end{Thm}
        \begin{prf}
            According to Lemma~\ref{lem:belief}, when $\delta \rightarrow 0$ and $i\rightarrow \infty$, $\bE_{\widehat{\boldsymbol \theta}_{i}} \bL_i \rightarrow \bar\bL$.
            Repeating the argument used in the proof of Theorem~\ref{thm:conv_0}, we finish the proof.\\
        \end{prf}
        
\section{Agent Influence and Explainability}
    \label{sec:explain}
    Now that we have devised a procedure to learn the graph topology, we can examine a useful question relating to how information flows over the network. Typically, there might exist more than one combination of edges leading from a node $\ell \in \mathcal N$ to some other node $k \in \mathcal N$. In the following, we describe one approach to measure path {\it influence} and find the most influential path between two nodes based on the extracted information about the graph $A_\star$. 
    
    According to the combination step~(\ref{eq:combine}), the combination weight $a_{\ell k} \neq 0$ of the edge linking node $\ell$ to $k$ reflects the direct (one-hop) influence of node $\ell$ on node $k$ for the belief construction. We can generalize the notion of the one-hop influence to any chosen path of length $r \geq 0$. We say that a sequence of edges $\left((\ell, v_1),\dots,(v_j, v_{j+1}),\dots,(v_r, k)\right)$ defines a path from $\ell$ to $k$ if the corresponding product of combination weights $a_{\ell v_1}\cdot\dots\cdot a_{v_rk}$ is positive. In other words, if these edges are present in the network.
    
    It is reasonable to expect that the influence of the path $\left((\ell, v_1),\dots,(v_r,k)\right)$ on the flow of information from $\ell$ to $k$ is proportional to the product of the combination weights along the path:
    \begin{align}
        \prod_{(\ell',k')\in \left((\ell, v_1),\dots,(v_r,k)\right)} a_{\ell',k'}.
        \label{eq:paths_prod}
    \end{align}
    We justify this statement as follows. 
    
    One popular technique in problems related to \textit{explainability} and \textit{interpretability} is sensitivity analysis~\cite{baehrens2010explain, adadi2018peeking}. We adapt this approach to our graph model. We treat the received observations $\boldsymbol\zeta_{k,i}$ at each individual agent $k \in \mathcal N$ as data available to the network. One way to measure the influence of a node $\ell \in \mathcal N$ on some other node $k \in \mathcal N$ is to assess how the belief vector $\boldsymbol{\psi}_{k,i}$ changes in response to variations at location $\ell$. To do so, we find it convenient to examine how the entries of $\blambda_i$ change with $\bL_t$. Thus, consider the following partial derivative expression (derived in Appendix~\ref{app:path}):
    \begin{align}
        \frac{\partial[\blambda_i]_{k,j}}{\partial[\bL_{t}]_{\ell, j}} = \delta (1-\delta)^{i-t} \sum_{((\ell, v_1),\dots,(v_{i-t-1}, k))} a_{\ell,v_1}\dots a_{v_{i-t-1},k}.
        \label{eq:part_derivative}
    \end{align}
    This expression is measuring how entries on the $k-$th row of $\blambda_i$ (which depend on $\boldsymbol{\psi}_{k,i}$) vary in response to changes in the entries on the $\ell-$th row of $\bL_t$ for any $t\leq i$ (which depend on the data $\boldsymbol{\zeta}_{\ell,t}$). To  avoid confusion, we clarify the case when $i=t$:
    \begin{align}
        \frac{\partial [\blambda_i]_{k,j}}{\partial [\bL_{i}]_{\ell, j}} = \begin{cases} \delta a_{\ell, \ell},& \textrm{ if } \ell = k,\nonumber\\
        0, & \textrm{ otherwise}.
        \end{cases}
    \end{align}
    Observe that the derivative in~(\ref{eq:part_derivative}) depends only on the time difference $i-t$, and the observations $\blambda_i$, $\bL_i$ are considered fixed. Result~(\ref{eq:part_derivative}) accommodates all paths of length $i-t$ leading from $\ell$ to $k$ and penalizes each one of them by $(1-\delta)^{i-t}$. Since combination weights lie in $[0,1]$, and the step size $0< \delta <1$, the influence drops as the length of the path grows. Thus, to measure the influence of $\ell$ on $k$, we will limit the path length by some hyperparameter $d$. It is reasonable to consider $d$ larger or equal to the shortest path length from node $\ell$ to $k$, or the average shortest path between all nodes in the network. In the next section, we will empirically confirm that a relatively small $d$ is enough because the influence of past data reduces with time drastically. 
    
    Motivated by ~(\ref{eq:part_derivative}), we will employ the following influence measure between the agents $\ell$ and $k$:
    \begin{align}
        &\;\eta_d(\ell,k) \triangleq \sum_{j = 1}^{|\Theta|-1} \sum_{r=0}^d \frac{\partial[\blambda_d]_{k,j}}{\partial[\bL_r]_{\ell, j}} \nonumber\\
        =&\; (|\Theta|-1) \delta \sum_{r=0}^{d} (1-\delta)^{r} \sum_{v_1,\dots,v_{r-1} \in \mathcal N} a_{\ell,v_1}\dots a_{v_{r-1},k}.
        \label{eq:node_influence_hop}
    \end{align}
    We derive the expression for $\eta_d(\ell,k)$ 
    in Appendix~\ref{apx:influences}. As a result, we observe that the influence of agent $\ell$ on agent $k$ is proportional to the sum of the products~(\ref{eq:paths_prod}) of paths leading from node $\ell$ to $k$. We therefore associate with each path from $\ell$ to $k$ an influence measure that depends on the product of the weights over the path and a second factor that depends on the length of the path, namely, 
    \begin{align}
        &\mathbb I \big(\left((\ell, v_1),\dots,(v_r,k)\right)\big) \nonumber \\
        \triangleq&\; (|\Theta|-1)\delta(1-\delta)^r \prod_{a_{\ell',k'}\in \left((\ell, v_1),\dots,(v_r,k)\right)} a_{\ell',k'},
        \label{eq:path_influence}
    \end{align}
    so that
    \begin{align}
        \eta_d(\ell,k) = \sum_{p \in \mathcal P^d_{\ell,k}} \mathbb I\left(p\right),
        \label{eq:path_influence_sum}
    \end{align}
    where $\mathcal P^d_{\ell, k}$ is the set of all paths leading from node $\ell$ to node $k$ with the lengths of these paths less than or equal to $d$.
    
    The influence formulation in terms of individual paths sheds light on how information flows in the network. First, the influences are not necessarily symmetric, $\eta_d(\ell,k)\neq\eta_d(k,\ell)$, since the combination matrix $A_\star$ is assumed to be only left-stochastic. This observation  provides an opportunity to explore in the future the question of \textit{causality}~\cite{michotte2017perception} and to examine questions related to the causal effect on node $k$. 
    
    The next question we discuss is how to identify the most  significant paths from $\ell$ to $k$, denoted by $p_{\ell, k}^{\star}$. To do so, we  search for the path that contributes the most to $\eta_d(\ell,k)$ -- see Appendix~\ref{app:path}:
    \begin{align}
        p_{\ell,k}^\star =&\; \arg\max_{p\in \mathcal P^d_{\ell,k}} \left\{ \mathbb I\left(p\right) \right\} \nonumber\\
        =&\; \arg\min_{p\in \mathcal P^d_{\ell,k}} \Bigg\{ - \sum_{(\ell'k')\in p} \big(\log a_{\ell'k'} + \log\left(1-\delta\right)\big) \Bigg\}.
    \end{align}
    In practice, this optimization problem can be solved using the Dijkstra algorithm~\cite{dijkstra1959note, chen2003dijkstra} for searching for the shortest path between nodes over a weighted graph, taking $-\log(a_{\ell'k'}) - \log(1-\delta)$ as edge weights in places where $a_{\ell'k'}$ is positive.

\section{Computer Simulations}
\label{sec:experiments}

    The experiments that follow help illustrate the ability of the proposed algorithm to identify edges and to adapt to situations where the graph topology is dynamic, as well as the hypothesis.
    
    \subsection{Setup}
        We consider a network of $30$ agents with $|\Theta|=10$ states, where the adjacency matrix is generated according to the Erdos-Renyi model with edge probability $p=0.2$. We set $\mathcal Z_k$ to be a discrete sample space with $|\mathcal Z_k| = 2$ for $k\in\mathcal N$. The step-size of the model is set to $\delta=0.1$. We define the likelihood functions $L_k(\theta)$, $k\in\mathcal N$, $\theta\in\Theta$, as follows:
        \begin{align}
            &L_k(\bm\zeta|\theta) = \sum_{z \in \mathcal Z_k}\mathbb I[\bm\zeta = z]\beta_{k,z}(\theta),\nonumber\\
            &\beta_{k,z}(\theta) \geq 0, z \in \mathcal Z_k\nonumber\\
            &\sum_{z \in \mathcal Z_k}\beta_{k,z}(\theta) = 1,
        \end{align}
        where the parameters $\beta_{k,z}(\theta)$ are generated randomly. During the graph learning procedure, we use $\mu=0.1$.
    
    \subsection{Graph learning}
\begin{figure}
            \centering
            \begin{subfigure}[b]{0.45\textwidth}
                \centering
                \subcaption{True graph.}
                \label{fig:adj_true}
                \includegraphics[width=0.9\textwidth]{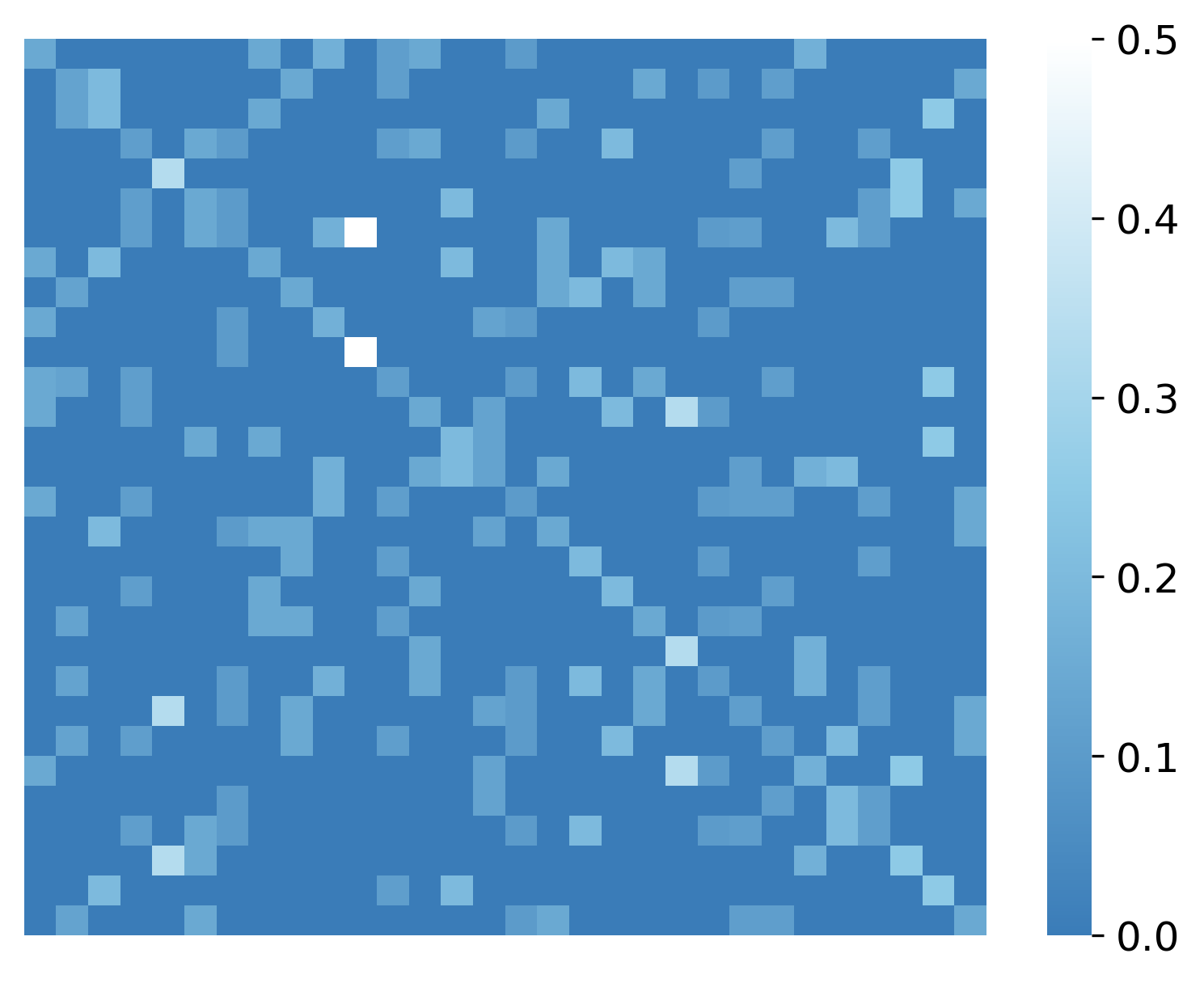}
            \end{subfigure}
            \vfill
            \begin{subfigure}[b]{0.45\textwidth}  
                \centering 
                \subcaption{Learned graph.}
                \label{fig:adj_learned}
                \includegraphics[width=0.9\textwidth]{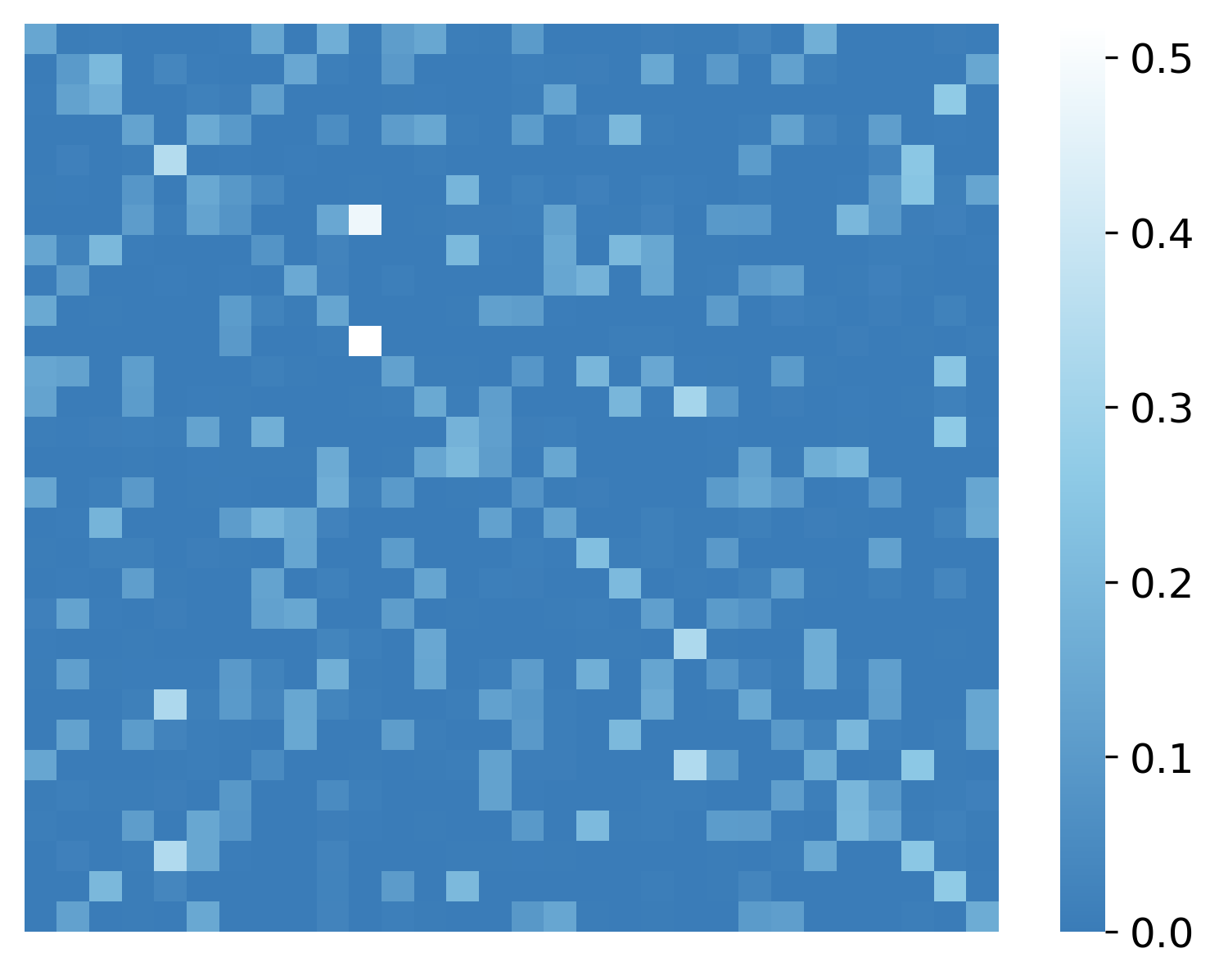}
            \end{subfigure}
            \caption{True combination matrix and the learned matrix using the Online Graph Learning (OGL) algorithm.}
            \label{fig:adj}
        \end{figure}

\begin{figure}
            \centering
            \begin{subfigure}[b]{0.4\textwidth}
                \centering
                \subcaption{True graph.}
                \label{fig:graph_true}
                \includegraphics[width=0.9\textwidth]{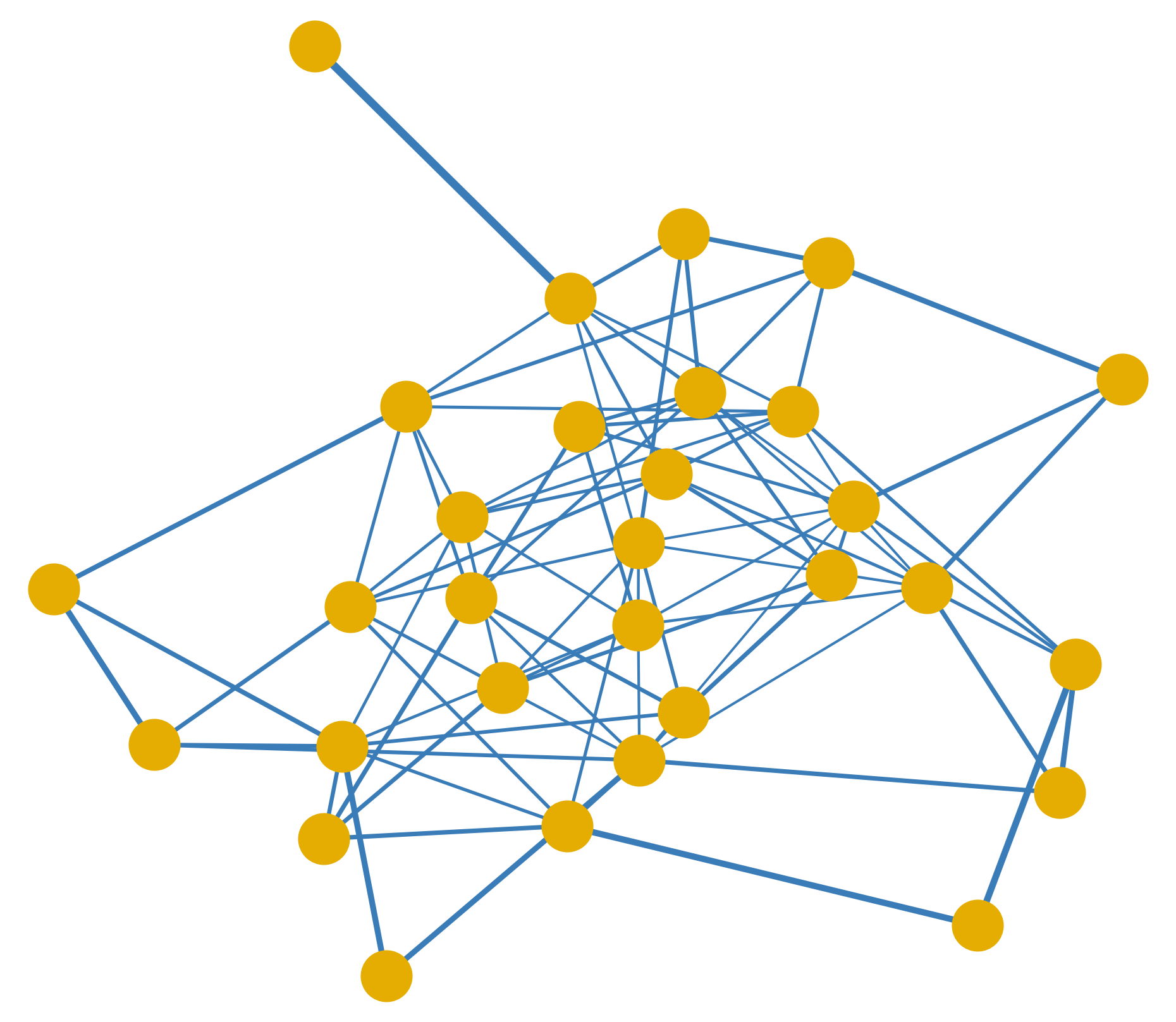}
            \end{subfigure}
            \vfill
            \begin{subfigure}[b]{0.4\textwidth}  
                \centering 
                \subcaption{Learned graph.}
                \label{fig:graph_learned}
                \includegraphics[width=0.9\textwidth]{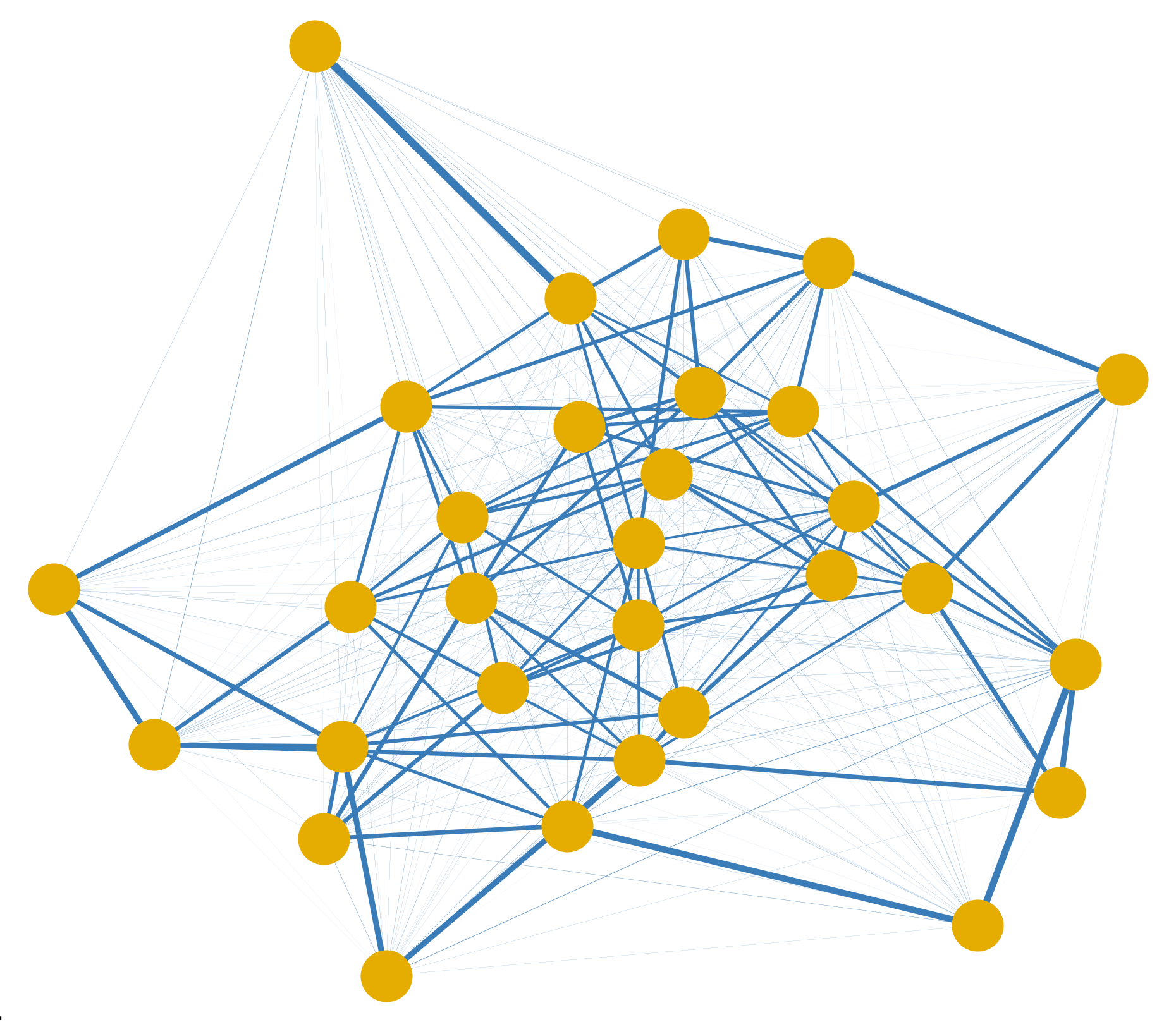}
            \end{subfigure}
            \caption{True combination graph and the learned graph using the Online Graph Learning (OGL) algorithm. Since graphs are bidirectional, each edge width is proportional to the sum of the weights on the edges.}
            \label{fig:graph}
        \end{figure}

        We provide a comparison between the true combination matrix and the estimated combination matrix. We plot the combination matrices in Fig.~\ref{fig:adj} and the graphs in Fig.~\ref{fig:graph}, where the width of an edge corresponds to the value of the combination weight. The experiment shows the ability of the algorithm to identify the graph: the recovered combination weights are close to the actual weights. In no-edge places, we observe reasonably small weights on the recovered matrix. These can be removed in post-processing by simple thresholding or more elaborate schemes, such as the $k$-means algorithm~\cite{matta2020graphlearning}.
        
\begin{figure}
            \centering
            \begin{subfigure}[b]{0.5\textwidth}
                \centering
                \subcaption{Algorithm performance when the true state changes.}
                \label{fig:state_change}
                \includegraphics[width=0.95\textwidth]{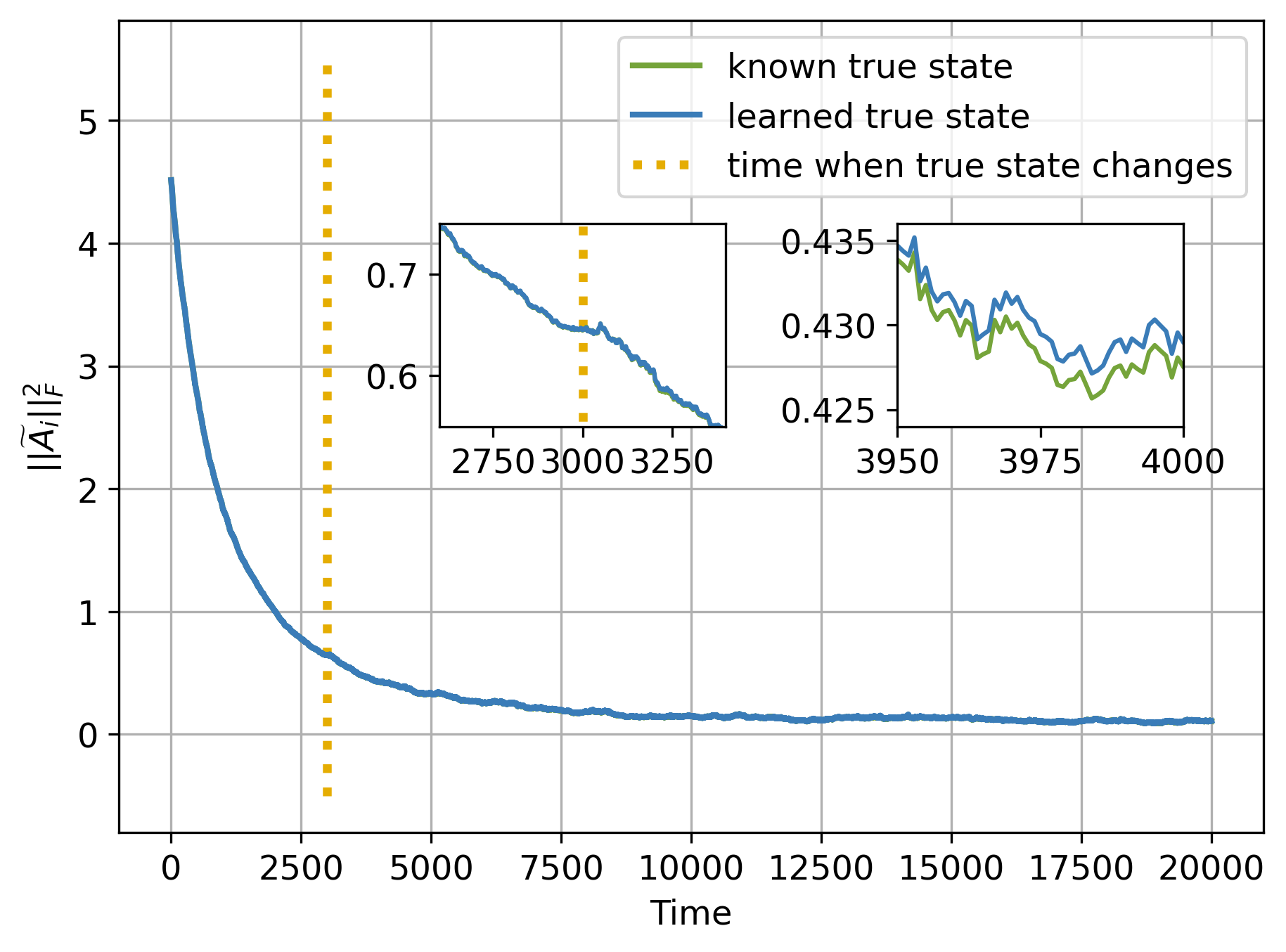}
            \end{subfigure}
            \vfill
            \begin{subfigure}[b]{0.5\textwidth}  
                \centering 
                \subcaption{Algorithm performance when the graph edges are regenerated.}
                \label{fig:graph_change}
                \includegraphics[width=0.95\textwidth]{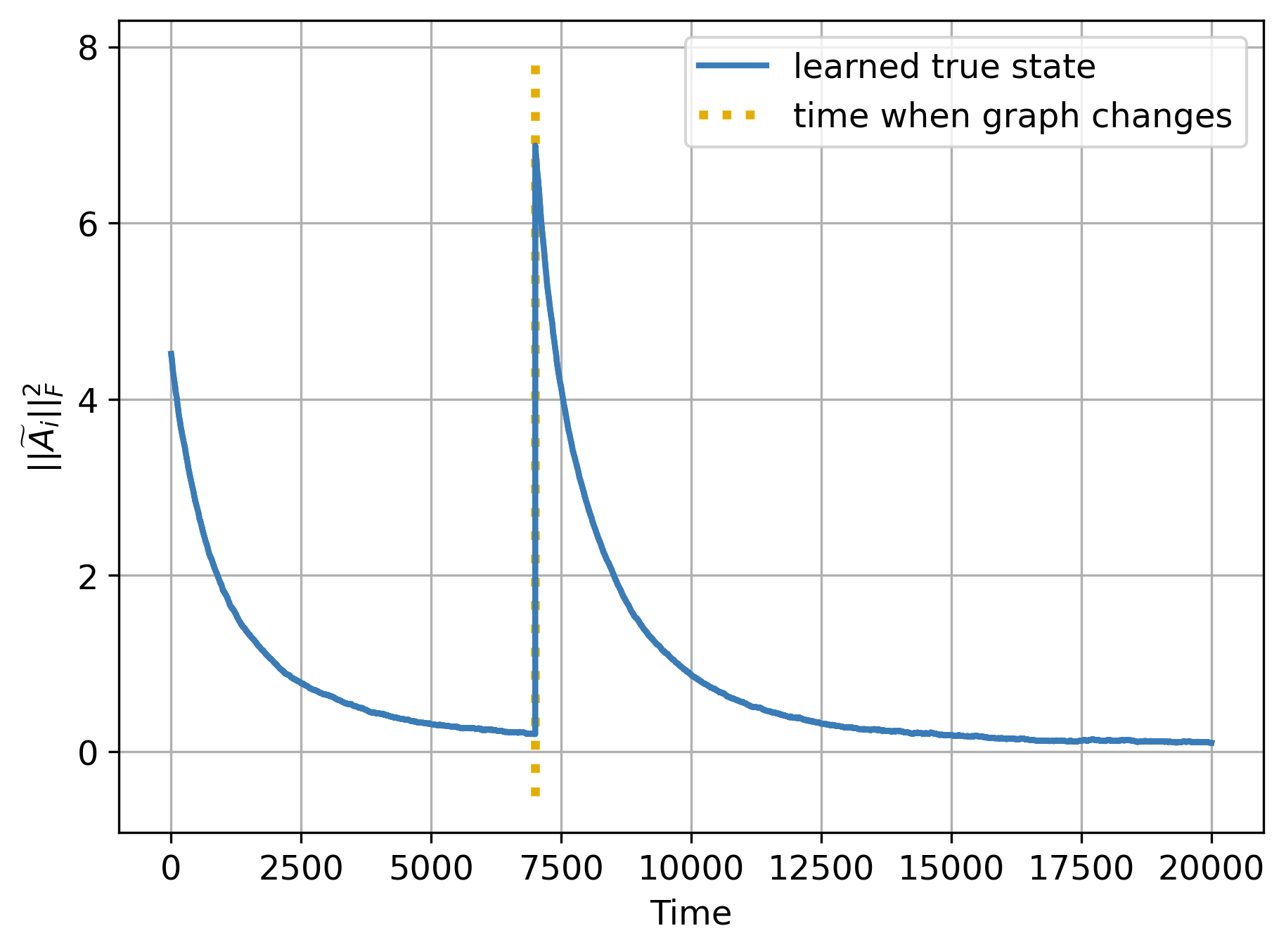}
            \end{subfigure}
            \vfill
            \begin{subfigure}[b]{0.5\textwidth}  
                \centering 
                \subcaption{Algorithm performance when the graph regularly changes. Every element of the adjacency matrix change its state with probability $0.5\%$.}
                \includegraphics[width=0.95\textwidth]{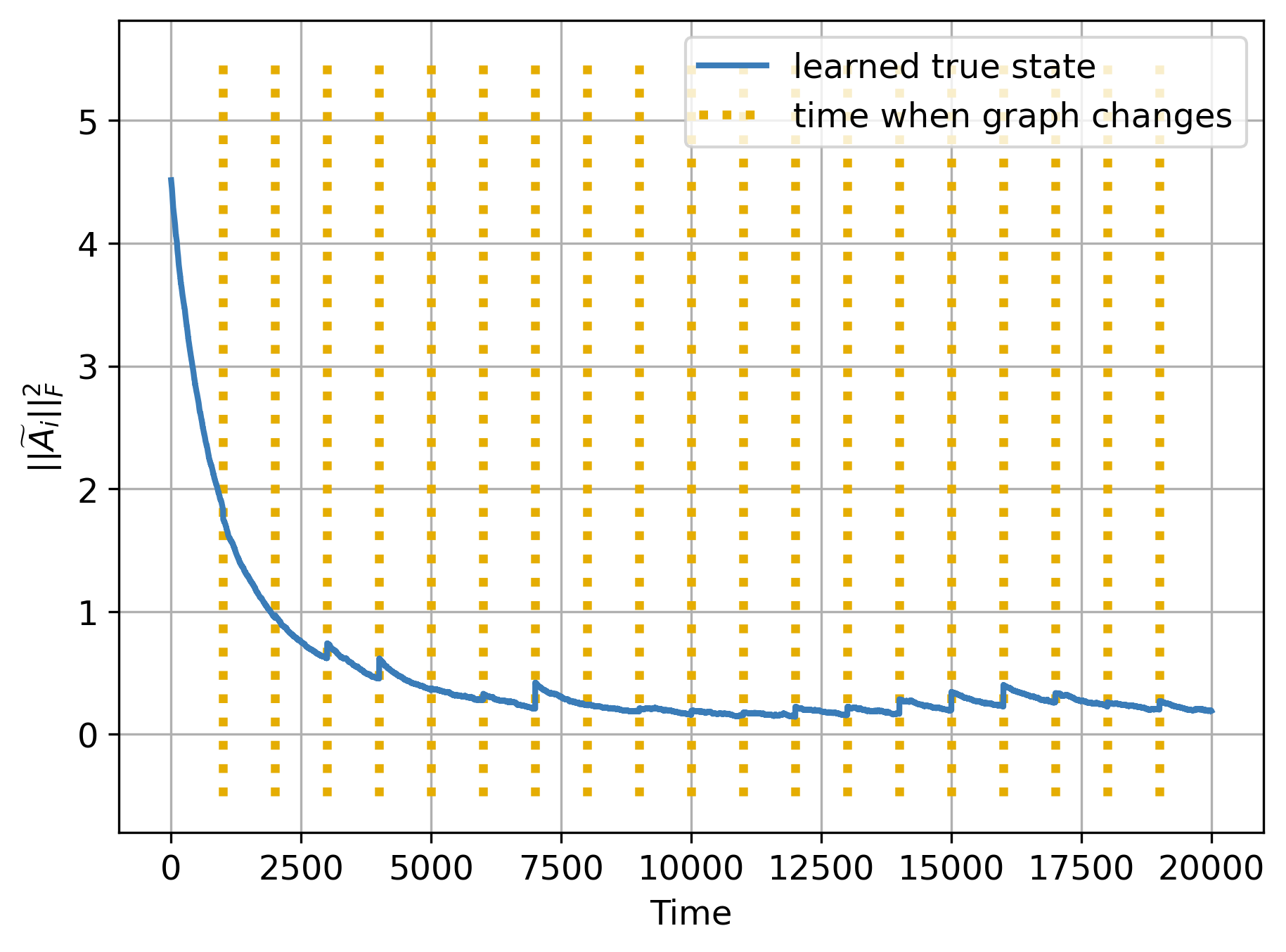}
                \label{fig:graph_change2}
            \end{subfigure}
            \caption{Algorithm performances.}
            \label{fig:errors}
        \end{figure}

        Additionally, in Fig.~\ref{fig:state_change} we plot how the deviation from the true matrix $A_\star$ evolves. The deviation is computed as the following quantity: 
        \begin{align}
            \|\widetilde{\bA}_i\|_{\rm F}^2 = \|A_\star - \bA_i\|_{\rm F}^2.
        \end{align} 
        We provide the error rates for both algorithm variants with known true state $\theta^\star$ and estimated true state $\widehat{\boldsymbol\theta}_i$. We see that there is a negligible gap in the learning performance.
        
        The proposed algorithm is robust to changes in the true state and graph topology. From Fig.~\ref{fig:state_change} we can see that changes to the true state have almost no impact on the learning performance. 
        In Fig.~\ref{fig:graph_change}, we regenerate edges at time 7000. The algorithm adapts and converges to the new combination matrix at a linear rate. A more challenging scenario is shown in Fig.~\ref{fig:graph_change2}, where we change the graph topology such that each edge changes (appears or disappears) with probability $0.5\%$ on a regular basis. Nevertheless, the graph learning algorithm is able to adapt and accurately recover the true combination matrix. Thus, we have experimentally illustrated that the algorithm is stable to dynamic network changes, which is a natural setting to consider in practice. These properties hold because the algorithm is online and processes data one by one with a constant learning rate $\mu>0$.
        
        \begin{Rem}
            The closest algorithm to compare with is the online algorithm for the heat diffusion process~\cite{vlaski2018online}. That work formulates the problem of graph identification based on streaming realizations of a vector signal $\boldsymbol s_i$. The recursion for the centered signal $\bar {\boldsymbol s}_i = \boldsymbol s_i - \frac 1 {|\mathcal N|} \mathds 1 \mathds 1^\bT$ obeys a recursion similar to~(\ref{eq:recursion_adaptive})
            \begin{align}
                \bar {\boldsymbol s}_i = W \bar {\boldsymbol s}_{i-1} + \bar {\boldsymbol p}_i,
            \end{align}
            and the combination matrix can be deduced from the matrix $W$. It is important to note that $\bE \bar {\boldsymbol p}_i = 0$, while in our model, $\bE \bL_i$ is unknown. This difference plays a significant role in our analysis. If we omit the assumption that $\theta^\star$ is unknown, our algorithm can be viewed as a variant of~\cite{vlaski2018online}. The experimental comparison of known and unknown true state variants is already provided in the current section.
        \end{Rem}
    
    \subsection{Agent Influence}
\begin{figure}
            \centering
            \includegraphics[width=0.4 \textwidth]{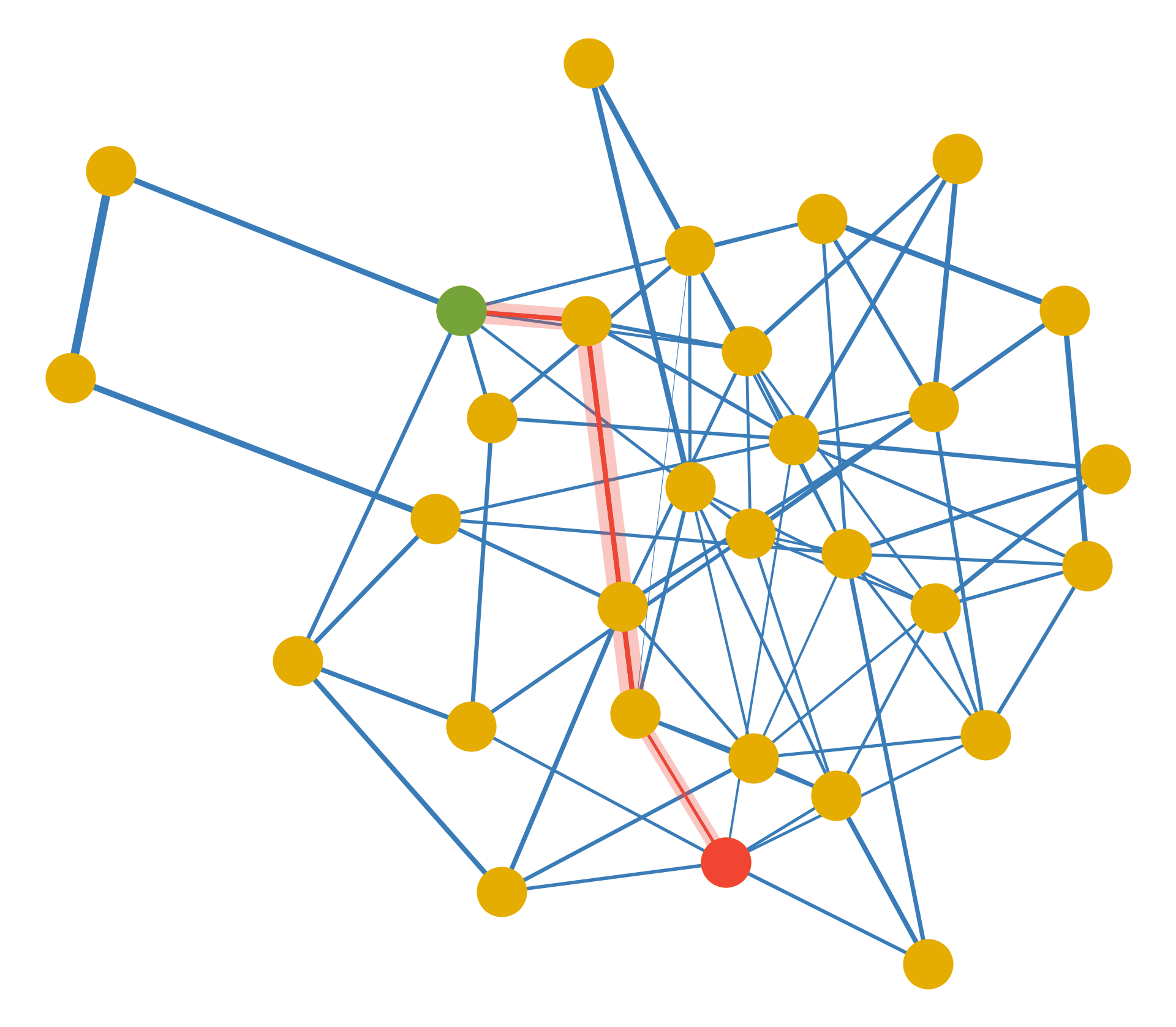}
            \caption{The red path illustrates the most influential path from the red node to the green node.}
            \label{fig:graph_path}
        \end{figure}

        The following experiments illustrate how to identify influences over the recovered graph. In Fig.~\ref{fig:graph_path}, we search for the most influential path $p_{\ell,k}^\star$. We observe that the algorithm finds the shortest path with the densest edges.
        
\begin{figure*}
        \centering
        \begin{subfigure}[b]{0.3\textwidth}
            \caption{$d=1$}
            \includegraphics[width=1\linewidth]{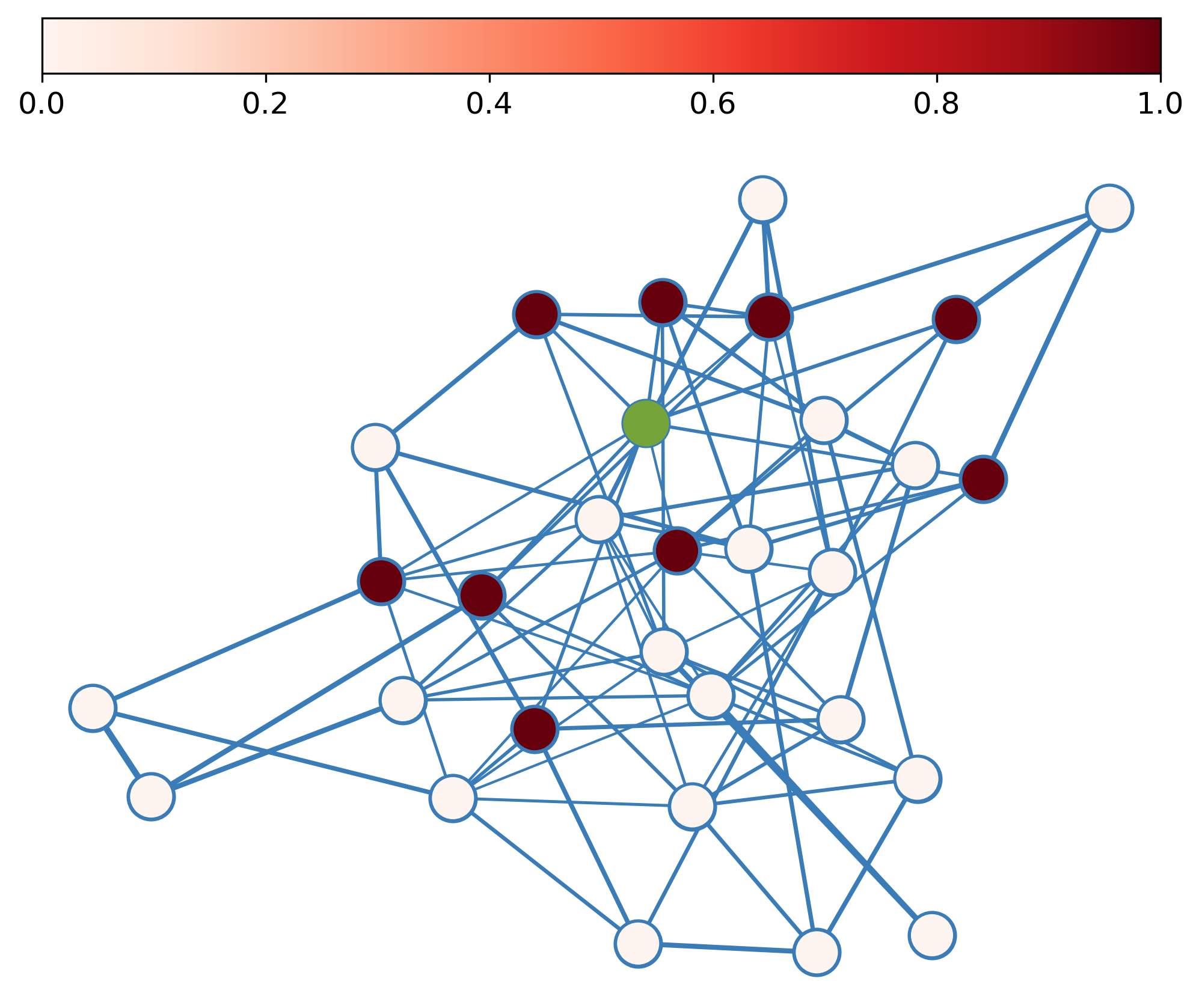}
            \label{fig:graph_path2_2}
        \end{subfigure}
        \begin{subfigure}[b]{0.3\textwidth}
            \caption{$d=2$}
            \includegraphics[width=1\linewidth]{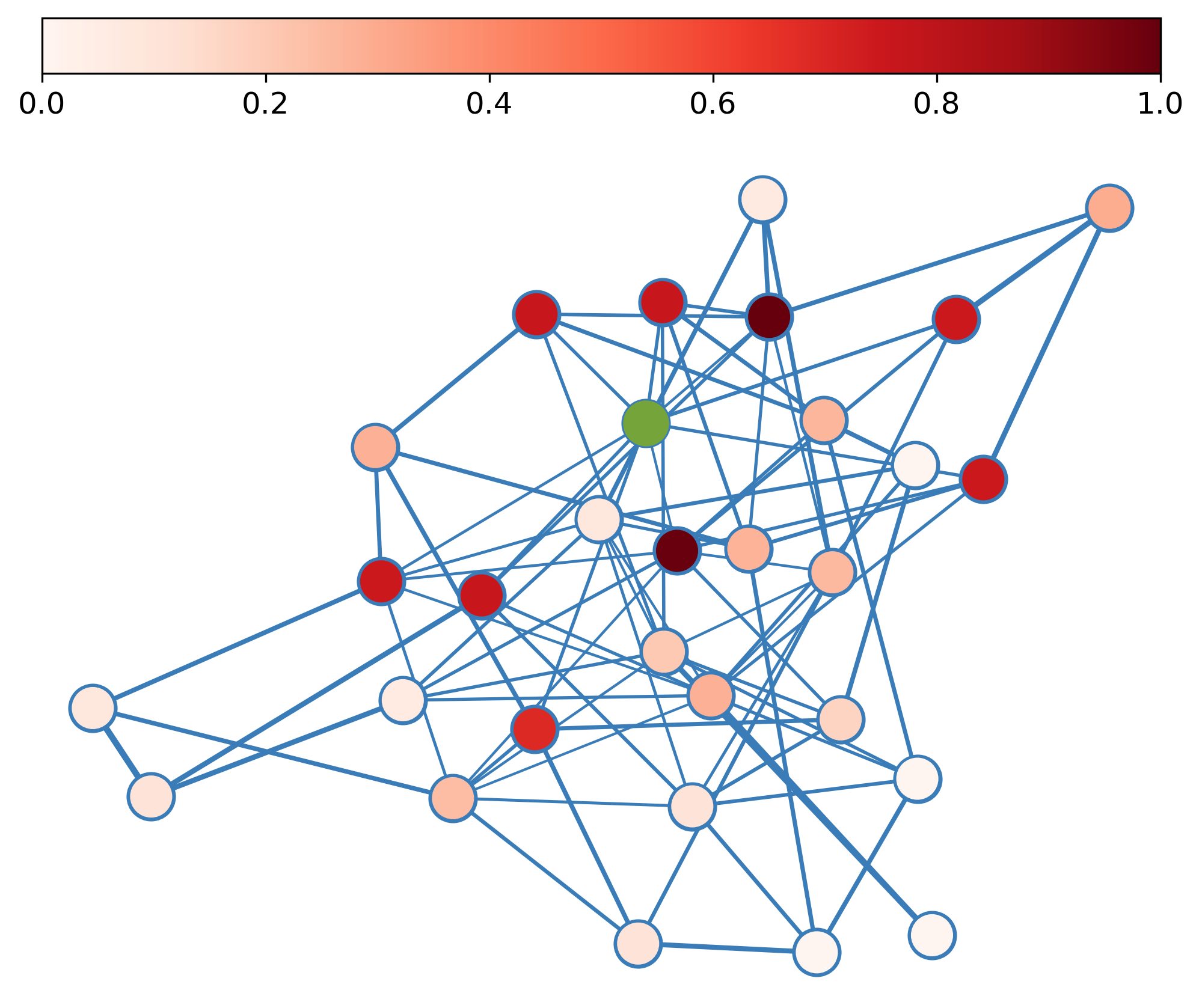}
            \label{fig:graph_path2_3}
        \end{subfigure}
        \begin{subfigure}[b]{0.3\textwidth}
            \caption{$d=3$}
            \includegraphics[width=1\linewidth]{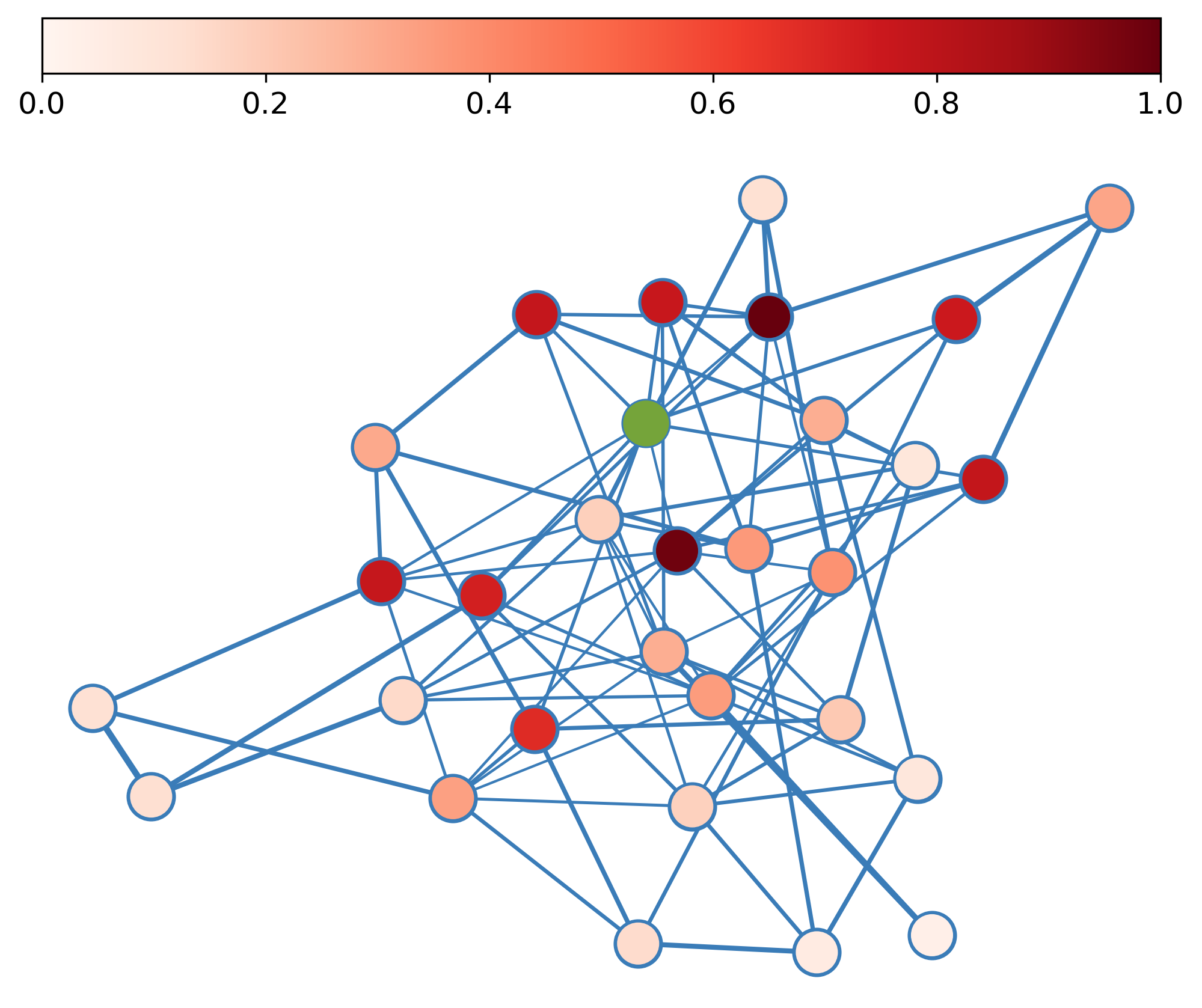}
            \label{fig:graph_path2_4}
        \end{subfigure}
        
        \caption{For different choice of $d$, the plots illustrate influences $\eta_d(\ell,k)$ on the target green node by all other agents in the network. The color intensity reflects the value of influence. The influences are normalized to $[0,1]$.}
        \label{fig:graph_path2}
        \end{figure*}

\begin{figure}
            \centering
            \includegraphics[width=0.4\textwidth]{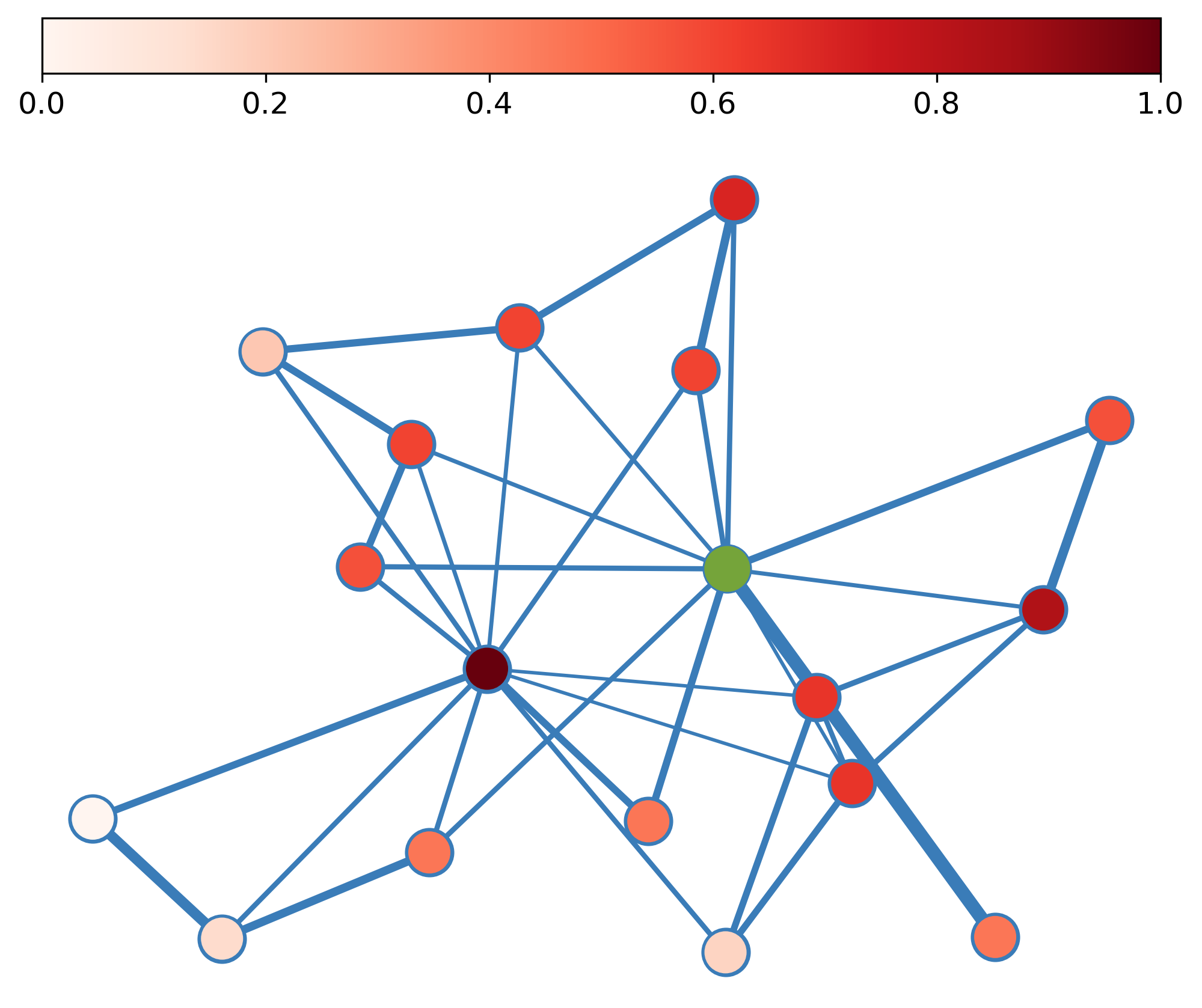}
            \caption{ The plot shows influences $\eta_3(\ell,k)$ on the target green node by all other agents in the network. The color intensity reflects the value of influence. The influences are normalized to $[0,1]$. In this example, a non-direct connection has more influence on the target node than the neighbors.}
            \label{fig:graph_star}
        \end{figure}

        In the next series of experiments shown in Fig.~\ref{fig:graph_path2}, we illustrate how the influences $\eta_d(\ell,k)$ are distributed for a fixed node $k \in \mathcal N$ for different $d$. As we already noted in the previous section, $\eta_d(\ell,k)$ sums influences along all paths of length from $0$ to $d$. Thus, we expect a relatively small $d$ to be enough for practical reasons. We experimentally verify that $d=2$ is optimal for our size graph since Figs.~\ref{fig:graph_path2_3} and~\ref{fig:graph_path2_4} show a very similar behavior. In the experiment, however, the network is relatively dense, and the nearest neighbors become the most influential ones. Fig.~\ref{fig:graph_star} illustrates a case where the most influential node is not a direct neighbor.
        
\section{Conclusions}
    In this paper, the problem of graph learning through observing social interactions by means of an adaptive social learning protocol is investigated. We develop an online algorithm that learns the agents' influence pattern via observing agents' beliefs over time. We prove that the proposed algorithm successfully learns the underlying combination weights and demonstrate its performance through analysis and computer simulations. In this way, we are able to discover the pattern of information flow in the network. A distinct feature of the proposed algorithm is the fact that it can track changes in the graph topology as well as in the true hypothesis.
    
    As future work, we may investigate the {\em partial information} setting, where the algorithm has access only to the beliefs of a subset of the network agents.

\appendices

\section{}\label{app:recursion}
    Consider a generic entry of $[\blambda_i]_{k,j}$. In view of~(\ref{eq:adapt_adaptive})--(\ref{eq:combine}), we can write:
    \begin{align}
        &[\blambda_i]_{k,j} = \log \frac{\bm\psi_{k,i}(\theta_j)}{\bm\psi_{k,i}(\theta_0)} = \log \frac{L_k^\delta(\boldsymbol{\zeta}_{k,i}|\theta_j)\boldsymbol{\mu}^{1-\delta}_{k,i-1}(\theta_j)}
        {L_k^\delta(\boldsymbol{\zeta}_{k,i}|\theta_0)\boldsymbol{\mu}^{1-\delta}_{k,i-1}(\theta_0)} \nonumber\\
        =&\; \log \frac{L_k^\delta(\bm\zeta_{k,i}|\theta_j)\left(\prod_{\ell}\bm\psi_{\ell,i-1}^{a_{\ell k}}(\theta_j)\right)^{1-\delta}}
        {L_k^\delta(\bm\zeta_{k,i}|\theta_0)\left(\prod_{\ell}\bm\psi_{\ell,i-1}^{a_{\ell k}}(\theta_0)\right)^{1-\delta}} \nonumber\\
        =&\; (1-\delta) \sum_{\ell} a_{\ell k} \log \frac{\bm\psi_{\ell,i-1}(\theta_j)}{\bm\psi_{\ell,i-1}(\theta_0)} +\textrm{ }\delta \log \frac{L_k(\bm\zeta_{k,i}|\theta_j)}{L_k(\bm\zeta_{k,i}|\theta_0)} \nonumber\\
        =&\; (1-\delta) \sum_{\ell} [A_\star]_{\ell,k} [\blambda_{i-1}]_{\ell, j} + \delta [\bL_i]_{k,j}
    \end{align}

\section{Proof of Lemma~\ref{lem:lambda}}\label{l1}
    \noindent Iterating~(\ref{eq:recursion_adaptive}) we get:
    \begin{align}
        \blambda_i = \left(1-\delta\right)^i \left(A_\star^i\right)^\bT\Lambda_0 + \delta \sum_{t=0}^{i-1} \left(1-\delta\right)^{t}(A_\star^{t})^\bT\bL_{i-t},
        \label{eq:recursion_extended}
    \end{align}
    where the entries of the initial matrix $\Lambda_0$ are given by:
    \begin{align}
        [\Lambda_0]_{k,j} = \log \frac{\boldsymbol\psi_{k,0}(\theta_0)}{\boldsymbol\psi_{k,0}(\theta_j)}.
    \end{align}
    The first term in~(\ref{eq:recursion_extended}) dies out as $i\rightarrow \infty$ since the eigenvalues of $A_{\star}$ are bounded by one in magnitude. Thus, in distribution, $\blambda_i$ converges to the following random matrix:
    \begin{align}
        &\blambda \triangleq \lim_{i\rightarrow\infty}\delta \sum_{t=0}^{i-1} \left(1-\delta\right)^{t}(A_\star^{t})^\bT\bL_{i-t} \nonumber\\
        \overset{\mathrm{d}}{=}& \lim_{i\rightarrow\infty}\delta \sum_{t=0}^{i-1} \left(1-\delta\right)^{t}(A_\star^{t})^\bT\bL_{t} = \delta \sum_{t=0}^\infty \left(1-\delta\right)^{t}(A_\star^{t})^\bT\bL_t,
    \end{align}
    where in the second equality we interchange $\bL_{t}$ by $\boldsymbol \bL_{i-t}$ because $\bL_i$ are i.i.d. and where the notation $\boldsymbol X\stackrel{\rm d}{=}\boldsymbol Y$ means that the variables  $\boldsymbol X$ and $\boldsymbol Y$ are equally distributed.
    
    Using Assumption~\ref{asm:support} we can upper-bound the elements of $|\blambda_i|$ as follows:
    \begin{align}
        [|\blambda_i|]_{k,j} \leq&\; \left(1-\delta\right)^i \sum_{\ell\in\mathcal N} [A_\star^i]_{\ell,k}[|\Lambda_0|]_{\ell,j} \nonumber\\
        &\;+ \delta \sum_{t=0}^{i-1} \left(1-\delta\right)^{t}\sum_{\ell\in\mathcal N}[A_\star^{t}]_{\ell,k}^\bT[|\bL_{i-t}|]_{\ell,j} \nonumber\\
        \leq &\; \left(1-\delta\right)^i \sum_{\ell\in\mathcal N} [A_\star^i]_{\ell,k}[|\Lambda_0|]_{\ell,j} \nonumber\\
        &\;+ \delta b \sum_{t=0}^{i-1} \left(1-\delta\right)^{t}\sum_{\ell\in\mathcal N}[A_\star^{t}]_{\ell,k}^\bT.
    \end{align}
    Therefore, we can write
    \begin{align}
        |\blambda_i| \preceq\;& \left(1-\delta\right)^i \left(A_\star^\bT\right)^i|\Lambda_0| + \delta b \sum_{t=0}^{i-1} \left(1-\delta\right)^{t}(A_\star^{t})^\bT \mathds{1} \mathds{1}^\bT\nonumber\\
        \triangleq \;& \bar\Lambda_i.
    \end{align}
    In the limit, $\bar \Lambda_i$ converges to:
    \begin{align}
        \bar\Lambda =\delta b \sum_{t=0}^\infty \left(1-\delta\right)^{t}(A_\star^{t})^\bT \mathds{1}\mathds{1}^\bT.
        \label{eq:upperbound}
    \end{align}
    which has bounded entries. Indeed, using Property~\ref{property} we have:
    \begin{align}
        [\bar\Lambda]_{k,j} =&\; b \delta \sum_{t=0}^\infty \left(1-\delta\right)^{t} \sum_{\ell\in\mathcal N} [{A_\star^t}]_{\ell, k} \nonumber\\
        \leq&\; b \delta \sum_{t=0}^\infty \left(1-\delta\right)^{t} \sum_{\ell\in\mathcal N} (u_{\ell} + \sigma\beta^t)\nonumber\\
        =&\; b \delta \sum_{t=0}^\infty \left(1-\delta\right)^{t} \sum_{\ell\in\mathcal N} u_{\ell} + b \delta \sum_{t=0}^\infty \left(1-\delta\right)^{t} \sum_{\ell\in\mathcal N} \sigma\beta^t \nonumber\\
        =&\; b \delta \sum_{t=0}^\infty \left(1-\delta\right)^{t} + b \delta \sigma |\mathcal N| \sum_{t=0}^\infty \left(\left(1-\delta\right)\beta\right)^{t} \nonumber\\
        =&\; b \left(1 + \frac{\delta \sigma|\mathcal N|}{1 - \beta(1-\delta)}\right).
    \end{align}
    Similarly, a lower bound for $[\bar\Lambda]_{k,j}$ is given by:
    \begin{align}
        [\bar\Lambda]_{k,j} \geq b \left(1 - \frac{\delta \sigma|\mathcal N|}{1 - \beta(1-\delta)}\right)
    \end{align}
    and we conclude that $\bar\Lambda$ has bounded entries.

\section{Proof of Lemma~\ref{lem:cost}}\label{l2}
    Assuming that technical conditions from the dominated convergence theorem are met, we exchange the expectation and gradient operations (see~\cite[ch. 3]{Sayed_2023}). In that case, the gradient of $J_i(A)$ relative to $A$ is given by:
    \begin{align}
        &\left(\nabla J_i(A)\right)^\bT = -(1-\delta) \nonumber\\
        &\;\;\;\;\;\;\;\;\times \bE \left(\blambda_i -  (1-\delta)A^\bT\blambda_{i-1} - \delta \bar\bL\right) \blambda_{i-1}^\bT.
    \end{align}

    For any matrices $A_1$, $A_2$:
    \begin{align}
        &\mathrm{Tr}\left(\left(\nabla J_i(A_1) -\nabla J_i(A_2)\right)^\bT\left(A_1 - A_2\right)\right) \nonumber\\
        =\;\;& (1-\delta)^2\mathrm{Tr}\left((A_1 - A_2)^\bT \bE \blambda_{i-1}\blambda_{i-1}^\bT (A_1-A_2)\right) \nonumber\\
        \leq\;\;& (1-\delta)^2 \lambda_{\max}(\bE\blambda_{i-1} \blambda_{i-1}^\bT) \|A_1 - A_2\|^2_{\mathrm{F}}.
        \label{eq:helper34}
    \end{align}
    In Appendix~\ref{appendix_lem} we show that $\bE\blambda_i\blambda_i^\bT$ and its limit are finite positive semi-definite matrices, and therefore their eigenvalues are finite. Result (\ref{eq:helper34}) justifies (\ref{eq:kappa}). Likewise, we can establish a lower bound with $\lambda_{\max}$ replaced by $\lambda_{\min}$ and arrive at (\ref{eq:nu}). 
    
    Next, by evaluating the gradient of $J_i(A)$ at $A_{\star}$ we find:
    \begin{align}
        \nabla \left(J_i(A_\star)\right)^\bT &= -(1-\delta)\bE(\blambda_i - (1-\delta)A_\star^\bT\blambda_{i-1} - \delta \bar\bL)\blambda_{i-1}^\bT \nonumber\\
        &= -(1-\delta)\delta (\bar\bL-\bar\bL)\bE\blambda_{i-1}^\bT = 0.
    \end{align}
    where we used (\ref{eq:recursion_adaptive}) and the fact that $\bL_i$ are independently distributed w.r.t. time $i$ due to i.i.d. local observations $\boldsymbol\zeta_{k,i}$. We conclude that $A_{\star}$ is a minimizer for $J_i(A)$. Since $J_i(A)$ is strictly convex, $A_\star$ is the unique minimizer.
    
\section{}\label{appendix_lem}
    It is obvious from~(\ref{eq:recursion_extended}) that
    \begin{align}
        &\lim_{i\rightarrow\infty} \bE\blambda_i= \delta\sum_{t=0}^\infty (1-\delta)^t (A_\star^t)^{\bT} \bar\bL ,
    \end{align}
    whereas we verify next that
    \begin{align}
        &\lim_{i\rightarrow\infty} \bE\blambda_i\blambda_i^\bT =\delta^2 \sum_{t=0}^\infty \left(1-\delta\right)^{2t}(A_\star^t)^\bT\bE \bL_t\bL_t^\bT A_\star^{t} \nonumber\\
        &\;\;\;\;+ \delta^2 \sum_{t_1, t_2=0, t_1\neq t_2}^\infty \left(1-\delta\right)^{t_1+t_2}(A_\star^{t_1})^\bT\bar\bL \bar \bL^\bT A_\star^{t_2}.
        \label{eq:lamlam_T0}
    \end{align}
    is a finite positive-definite matrix with
    \begin{align}
        \lim_{i\rightarrow\infty} \bE\blambda_i\blambda_i^\bT \succeq \tau \delta^2 I.
    \end{align}
    
    \begin{prf}
        Note first, with an appropriate change of variables, that 
        \begin{align}
            &\blambda_i\blambda_i^\bT \nonumber\\
            =\textrm{ }& (1-\delta)^{2i} \left(A_\star^i\right)^\bT \Lambda_0\Lambda_0^\bT A_\star^i \nonumber\\
            \textrm{ }& + \delta^2 \sum_{t=1}^i \left(1-\delta\right)^{2i-2t}(A_\star^{i-t})^\bT\bL_t\bL_t^\bT A_\star^{i-t} \nonumber\\
            \textrm{ }& + \delta^2 \sum_{t_1, t_2=1, t_1\neq t_2}^i \left(1-\delta\right)^{2i-t_1-t_2}(A_\star^{i-t_1})^\bT\bL_{t_1}\bL_{t_2}^\bT A_\star^{i-t_2} \nonumber\\
            \textrm{ }& + \delta (1-\delta)^i (A_\star^i)^\bT \Lambda_0 \sum_{t=1}^i \left(1-\delta\right)^{i-t} \bL_t^\bT A_\star^{i-t} \nonumber \\
            \textrm{ }& + \delta (1-\delta)^i \left(\sum_{t=1}^i \left(1-\delta\right)^{i-t}(A_\star^{i-t_1})^\bT \bL_t\right)  \Lambda_0^\bT A_\star^i.  \nonumber \\
        \end{align}
        Taking expectation, we obtain:
        \begin{align}
            &\bE \blambda_i \blambda_i^\bT \nonumber\\
            =\textrm{ }& (1-\delta)^{2i} (A_\star^i)^\bT \Lambda_0\Lambda_0^\bT A_\star^i \nonumber\\
            \textrm{ }& + \delta^2 \sum_{t=0}^{i-1} \left(1-\delta\right)^{2t} (A_\star^t)^\bT\bE\bL_t\bL_t^\bT A_\star^{t} \nonumber\\
            \textrm{ }& + \delta^2 \sum_{t_1, t_2=0, t_1\neq t_2}^{i-1} \left(1-\delta\right)^{t_1+t_2}(A_\star^{t_1})^\bT\bar\bL \bar\bL^\bT A_\star^{t_2} \nonumber\\
            \textrm{ }& + \delta (1-\delta)^i (A_\star^i)^\bT \Lambda_0 \sum_{t=0}^{i-1} \left(1-\delta\right)^{t} \bar\bL^\bT A_\star^{t} \nonumber\\
            \textrm{ }& + \delta (1-\delta)^i \sum_{t=0}^{i-1} \left(1-\delta\right)^{i-t} (A_\star^t)^\bT \bar\bL \Lambda_0^\bT A_\star^i.
            \label{eq:lamlam_helper}
        \end{align}
        Therefore, in the limit:
        \begin{align}
            &\lim_{i\rightarrow\infty}\bE \blambda_i \blambda_i^\bT \nonumber\\
            =\textrm{ }& \delta^2 \sum_{t=0}^\infty \left(1-\delta\right)^{2t}(A_\star^t)^\bT\bE \bL_t\bL_t^\bT A_\star^{t} \nonumber\\
            \textrm{ }& + \delta^2 \sum_{t_1, t_2=0, t_1\neq t_2}^\infty \left(1-\delta\right)^{t_1+t_2}(A_\star^{t_1})^\bT\bar\bL \bar \bL^\bT A_\star^{t_2}.
        \end{align}
        Under Assumption~\ref{asm:support}, for any $k_1$, $k_2\in\mathcal N$:
        \begin{align}
            &\big|\bE [\bL_t\bL_t^\bT]_{k_1,k_2}\big| \leq \sum_j \big|\bE[\bL_t]_{k_1,j } [\bL_t]_{k_2,j}\big| \nonumber\\
            \leq&\; \begin{cases}
                \sum_j \bE[|\bL_t|]_{k_1,j} \bE[|\bL_t|]_{k_2,j}, \textrm{ if } k_1 \neq k_2,\\
                \sum_j \bE[|\bL_t|]^2_{k_1,j}, \;\;\;\;\;\;\;\;\;\;\;\;\;\;\;\;\textrm{ if } k_1 = k_2
            \end{cases} \nonumber\\
            \leq&\; b^2 (|\Theta|-1)
        \end{align}
        and
        \begin{align}
            \big|[\bar\bL\bar\bL^\bT]_{k_1,k_2}\big| \leq b^2(|\Theta|-1).
        \end{align}
        Thus, 
        \begin{align}
            &\big|\lim_{i\rightarrow\infty} \bE \blambda_i\blambda^\bT \big| \leq \delta^2 b^2 (1-|\Theta|) \nonumber\\
            &\; \times \sum_{t_1=0}^\infty\sum_{t_2=0}^\infty (1-\delta)^{t_1+t_2}(A_\star^\bT)^{t_1+t_2} \mathds 1 \mathds 1^\bT A_\star^{t_1+t_2} \nonumber\\
            &\; = \delta^2 b^2 (1-|\Theta|) \left[ \sum_{t=0}^\infty (1-\delta)^t(A_\star^\bT)^t \mathds 1 \mathds 1^\bT A_\star^t \right]^2
            \label{eq:lem_bounded}
        \end{align}
        In view of Property~\ref{property}, consider
        \begin{align}
            &[(A_\star^\bT)^t \mathds 1 \mathds 1^\bT A_\star^t]_{k_1,k_2} = [(A_\star^\bT)^t \mathds 1]_{k_1} [(A_\star^\bT)^t \mathds 1]_{k_2}\nonumber\\
            \leq&\; \sum_{\ell}[A_\star]_{\ell k_1} \sum_{\ell}[A_\star]_{\ell k_2} \leq \left[\sum_{\ell} \left(u_\ell + \sigma\beta^t\right)\right]^2 \nonumber\\
            =&\; \left(1 + \sigma \beta^t |\mathcal N|\right)^2.
        \end{align}
        Therefore, (\ref{eq:lem_bounded}) has bounded entries:
        \begin{align}
            &\big|\lim_{i\rightarrow\infty} \bE \blambda_i\blambda_i^\bT \big|_{k_1, k_2} \nonumber\\
            \leq&\; \delta^2 b^2 (1-|\Theta|) \left[\sum_{t=0}^\infty (1-\delta)^t \left(1 + \sigma \beta^t |\mathcal N|\right)^2 \right]^2 \nonumber\\
            =&\; \delta^2b^2(1-|\Theta|) \left[ \frac 1\delta + \frac{2\sigma|\mathcal N|}{1-\beta(1-\delta)} + \frac{\delta^2|\mathcal N|^2}{1-\beta^2(1-\delta)} \right]^2.
        \end{align}
        Next, let us show that $\lim_{i\rightarrow\infty} \bE \blambda_i\blambda_i^\bT$ is positive-definite. Under Assumption~\ref{asm:loglikelihoods_positivedefinite}, $\bE\bL_t \bL_t \succeq \tau I$ for any $t$. Consider the square root $B_t$, such that $\bE \bL_t\bL_t^\bT = B_t B_t^\bT$. Then,~(\ref{eq:lamlam_T0}) becomes
        \begin{align}
        &\lim_{i\rightarrow\infty} \bE\blambda_i\blambda_i^\bT =\delta^2\bE \bL_0\bL_0^\bT \nonumber\\
        &\;\;\;\;+ \delta^2 \sum_{t=1}^\infty \left(1-\delta\right)^{2t}( B_t^\bT A_\star^t )^\bT B_t^\bT A_\star^{t} \nonumber\\
        &\;\;\;\;+ \delta^2 \sum_{t_1, t_2=0, t_1\neq t_2}^\infty \left(1-\delta\right)^{t_1+t_2}(\bar\bL ^\bT A_\star^{t_1})^\bT \bar \bL^\bT A_\star^{t_2}.
        \label{eq:lamlam_T}
        \end{align}
        The equation above is a sum of a positive-definite matrix $\delta^2 \bE \bL_0 \bL_0^\bT \succeq \tau \delta^2 I$ and positive semi-definite matrices. Therefore, 
        \begin{align}
            \lim_{i\rightarrow\infty} \bE\blambda_i\blambda_i^\bT \succeq \tau \delta^2 I.
        \end{align}
    \end{prf}
    
\section{Proof of Theorem~\ref{thm:conv_0}}\label{t1}
    Each step of the graph learning procedure~(\ref{eq:graph_update_unbiased}) has the following form:
    \begin{align}
        \boldsymbol{A}_i &=  \boldsymbol{A}_{i-1} + \mu(1-\delta)\blambda_{i-1}\nonumber\\
        \textrm{ }&\;\;\;\;\times\left(\blambda_i^{\mathsf{T}} -  (1-\delta)\blambda_{i-1}^{\mathsf{T}}\boldsymbol{A}_{i-1} - \delta \bar\bL^\bT\right)\nonumber\\
        \textrm{ }&=\boldsymbol{A}_{i-1} + \mu(1-\delta)\blambda_{i-1}\nonumber\\
        \textrm{ }&\;\;\;\;\times\Big(
        (1-\delta)\blambda_{i-1}^\bT A_\star + \delta\bL_i^\bT
        -  (1-\delta)\blambda_{i-1}^{\mathsf{T}}\boldsymbol{A}_{i-1}\nonumber\\
        \textrm{ }&\;\;\;\;\;\;\;\;\;\;- \delta \bar\bL^\bT\Big)\nonumber\\
        \textrm{ }&=
        \bA_{i-1} + \mu(1-\delta)^2\blambda_{i-1}\blambda_{i-1}^\bT \widetilde{\bA}_{i-1}\nonumber\\
        \textrm{ }&\;\;\;\;+\mu\delta(1-\delta)\blambda_{i-1}(\bL_i - \bar\bL)^\bT
        \label{eq:helper1}
    \end{align}
    where in the second equality we used (\ref{eq:recursion_adaptive}) and introduced 
    \begin{align}
        \widetilde{\bA}_{i-1} = A_\star - \bA_{i-1}.
    \end{align}
    Subtracting $A_{\star}$ from both sides we get
    \begin{align}
        \widetilde{\bA}_i =& \left(I - \mu\left(1-\delta\right)^2\blambda_{i-1}\blambda_{i-1}^\bT\right)\widetilde{\bA}_{i-1} \nonumber\\
        &- \mu\delta\left(1-\delta\right)\blambda_{i-1}(\bL_i - \bar{\bL})^\bT\nonumber\\
        =& \left(I - \mu \left(1-\delta\right)^2\bE  \big[\blambda_{i-1}\blambda_{i-1}^\bT\big]\right)\widetilde{\bA}_{i-1} \nonumber\\
        &+\mu(1-\delta)^2\left(\bE \big[\blambda_{i-1}\blambda_{i-1}^\bT\big] - \blambda_{i-1}\blambda_{i-1}^\bT\right)\widetilde{\bA}_{i-1}\nonumber\\
        &- \mu\delta\left(1-\delta\right)\blambda_{i-1}(\bL_i - \bar{\bL})^\bT,
    \end{align}
    where $I$ stands for the identity matrix. Next, using the squared  Frobenius norm, computing the conditional expectation relative to the filtration $\boldsymbol{\mathcal F}_{i-1} = \{\boldsymbol\zeta_{k, j}, j < i, k\in \mathcal N\}$, and appealing to the separation principle from Assumption~\ref{asm:independence} we get:
    \begin{align}
        &\bE\left[\|\widetilde{\bA}_i\|_{\rm F}^2 \big| \boldsymbol{\mathcal F}_{i-1}\right] \nonumber\\
        &\overset{(a)}{=}  \|\big(I - \mu(1-\delta)^2 \bE \big[\blambda_{i-1}\blambda_{i-1}^\bT\big] \big) \widetilde{\bA}_{i-1}\|_{\rm F}^2\nonumber\\
        &\;\;\;\; + \mu^2 \|(1-\delta)^2 \big(\bE \big[\blambda_{i-1}\blambda_{i-1}^\bT\big] - \blambda_{i-1}\blambda_{i-1}^\bT\big)\widetilde{\bA}_{i-1}\|_{\rm F}^2\nonumber\\
        &\;\;\;\;+ \mu^2 \bE \left[ \|\delta(1-\delta)\blambda_{i-1}(\bL_i - \bar{\bL})^\bT\|_{\rm F}^2\big| \boldsymbol{\mathcal F}_{i-1}\right]\nonumber\\
        &\;\;\;\; + 2 \mu (1-\delta)^2 \operatorname{Tr}\Big( \big(I - \mu(1-\delta)^2 \bE \big[\blambda_{i-1}\blambda_{i-1}^\bT\big] \big) \widetilde{\bA}_{i-1} \nonumber\\
        &\;\;\;\;\;\;\;\; \times\widetilde{\bA}_{i-1}^\bT \big(\bE \big[\blambda_{i-1}\blambda_{i-1}^\bT\big] - \blambda_{i-1}\blambda_{i-1}^\bT\big) \Big) \nonumber\\
        &\leq \rho^2 \left(I - \mu(1-\delta)^2\bE \big[\blambda_{i-1}\blambda_{i-1}^\bT\big] \right) \|\widetilde{\bA}_{i-1}\|_{\rm F}^2\nonumber\\
        &\;\;\;\; + \mu^2 \|(1-\delta)^2 \left(\bE \big[\blambda_{i-1}\blambda_{i-1}^\bT\big] - \blambda_{i-1}\blambda_{i-1}^\bT\right)\|_{\rm F}^2 \|\widetilde{\bA}_{i-1}\|_{\rm F}^2\nonumber\\
        &\;\;\;\;+ \mu^2 \bE \left[\|\delta(1-\delta)\blambda_{i-1}(\bL_i - \bar{\bL})^\bT\|_{\rm F}^2\big| \boldsymbol{\mathcal F}_{i-1}\right]\nonumber\\
        &\;\;\;\; + 2 \mu (1-\delta)^2 \operatorname{Tr}\Big( \big(I - \mu(1-\delta)^2 \bE \big[\blambda_{i-1}\blambda_{i-1}^\bT\big] \big) \widetilde{\bA}_{i-1} \nonumber\\
        &\;\;\;\;\;\;\;\; \times\widetilde{\bA}_{i-1}^\bT \big(\bE \big[\blambda_{i-1}\blambda_{i-1}^\bT\big] - \blambda_{i-1}\blambda_{i-1}^\bT\big) \Big),
        \label{eq:th1_rec_сonditional}
    \end{align}
    where $\rho(\cdot)$ refers to the spectral radius of its matrix argument. In (a), only one cross-term in the form of a trace remains, since the other terms are zero under conditional expectation. Denoting
    \begin{align}
        \alpha_{i-1} \triangleq\;\;& \rho^2 \left(I - \mu(1-\delta)^2\bE \big[\blambda_{i-1}\blambda_{i-1}^\bT\big] \right) \nonumber\\
        &+ \mu^2\|(1-\delta)^2 \big(\bE \big[\blambda_{i-1}\blambda_{i-1}^\bT\big] - \blambda_{i-1}\blambda_{i-1}^\bT\big)\|_{\rm F}^2,
        \label{eq:helper11}
    \end{align}
    we get
    \begin{align}
        &\bE\left[\|\widetilde{\bA}_i\|_{\rm F}^2 \big| \boldsymbol{\mathcal F}_{i-1}\right]
        \leq  \alpha_{i-1} \|\widetilde{\bA}_{i-1}\|_{\rm F}^2\nonumber\\
        &\;\;\;\;\;\;+ \mu^2 \bE \left[ \|\delta(1-\delta)\blambda_{i-1}(\bL_i - \bar{\bL})^\bT\|_{\rm F}^2\big| \boldsymbol{\mathcal F}_{i-1}\right] \nonumber\\
        &\;\;\;\;\;\;+ 2 \mu (1-\delta)^2 \operatorname{Tr}\Big( \big(I - \mu(1-\delta)^2 \bE \big[\blambda_{i-1}\blambda_{i-1}^\bT\big] \big) \widetilde{\bA}_{i-1} \nonumber\\
        &\;\;\;\;\;\;\;\;\;\; \times\widetilde{\bA}_{i-1}^\bT \big(\bE \big[\blambda_{i-1}\blambda_{i-1}^\bT\big] - \blambda_{i-1}\blambda_{i-1}^\bT\big) \Big).
        \label{eq:th1helper}
    \end{align}
    Consider:
    \begin{align}
        &\rho^2 \left(I - \mu(1-\delta)^2 \bE \blambda_{i-1}\blambda_{i-1}^\bT \right) \nonumber\\
        =&\; \max \Big\{\left(1 - \mu(1-\delta)^2 \lambda_{\min}\left(\bE \blambda_{i-1}\blambda_{i-1}^\bT\right) \right)^2,\nonumber\\
        &\;\;\;\;\;\;\;\;\;\;\;\left(1 - \mu(1-\delta)^2 \lambda_{\max}\left(\bE \blambda_{i-1}\blambda_{i-1}^\bT\right) \right)^2\Big\}\nonumber\\
        =&\; \max \{ (1-\mu \nu_i)^2, (1-\mu\kappa_i)^2 \} \leq 1 - 2\mu\nu_i + \mu^2\kappa_i^2.
    \end{align}
    second term in (\ref{eq:helper11}) can be bounded as
    \begin{align}
        &\mu^2\|(1-\delta)^2 \big(\bE \blambda_{i-1}\blambda_{i-1}^\bT - \blambda_{i-1}\blambda_{i-1}^\bT\big)\|_{\rm F}^2\nonumber\\
        &\leq 2\mu^2\|(1-\delta)^2 \bE \blambda_{i-1}\blambda_{i-1}^\bT\|_{\rm F}^2 + 2\mu^2\|(1-\delta)^2 \blambda_{i-1}\blambda_{i-1}^\bT\|_{\rm F}^2\nonumber\\
        &\leq 2\mu^2\|(1-\delta)^2 \bE \blambda_{i-1}\blambda_{i-1}^\bT\|_{\rm F}^2 + 2\mu^2\|(1-\delta)^2 \bar \Lambda_{i-1}\bar \Lambda_{i-1}^\bT\|_{\rm F}^2
    \end{align}
    where we use Lemma~\ref{lem:lambda}. To clarify, consider each squared element
    \begin{align}
        &[\blambda_{i-1}\blambda_{i-1}^\bT]_{k,\ell}^2 \nonumber\\
        =&\; \Big( \sum_j [\blambda_{i-1}]_{k,j} [\blambda_{i-1}]_{\ell, j} \Big)^2 \leq  \Big( \sum_j [|\blambda_{i-1}|]_{k,j} [|\blambda_{i-1}|]_{\ell, j} \Big)^2 \nonumber\\
        \leq&\;  \Big( \sum_j [\bar\Lambda_{i-1}]_{k,j} [\bar\Lambda_{i-1}]_{\ell, j} \Big)^2 = [\bar\Lambda_{i-1}\bar\Lambda_{i-1}^\bT]^2_{k,\ell}.
    \end{align}
    Thus,
    \begin{align}
        \alpha_{i-1} \leq &\; 1 - 2\mu\nu_i + \mu^2\kappa_i^2 + 2\mu^2\|(1-\delta)^2 \bE \blambda_{i-1}\blambda_{i-1}^\bT\|_{\rm F}^2 \nonumber\\
        &\;+ 2\mu^2\|(1-\delta)^2 \bar{\Lambda}_{i-1}\bar{\Lambda}_{i-1}^\bT\|_{\rm F}^2 \nonumber\\
        =&\; 1 - 2 \mu\nu_i + O(\mu^2).
    \end{align}
    When $\mu$ is small enough, $\alpha_{i-1}$ is bounded away from one.
    For simplicity of notations, consider next
    \begin{align}
        \alpha_{i-1} \triangleq 1 - 2 \mu\nu_i + O(\mu^2).
        \label{eq:alpha_helper}
    \end{align}
    Returning to the other remaining terms in~(\ref{eq:th1helper}) we have:
    \begin{align}
        &\mu^2 \bE \left[ \|\delta(1-\delta)\blambda_{i-1}(\bL_i - \bar{\bL})^\bT\|_{\rm F}^2\big| \boldsymbol{\mathcal F}_{i-1}\right]\nonumber\\
        &=\mu^2 \delta^2(1-\delta)^2 \mathrm{Tr} \big( \blambda_{i-1} \R_{\bL} \blambda_{i-1}^\bT \big),
        \label{eq:th1helper2}
    \end{align}
    where $\R_{\bL} = \bE \left(\bL_i - \bar\bL\right)\left(\bL_i - \bar\bL\right)^\bT$ is independent of $i$ since $\mathcal L_i$ are i.i.d.
    By the law of total expectations, we rewrite~(\ref{eq:th1helper}) using~(\ref{eq:th1helper2}) and the fact that the last cross-term has zero mean:
    \begin{align}
        \bE\|\widetilde{\bA}_i\|_{\rm F}^2 \leq\;\; &\alpha_{i-1} \bE \|\widetilde{\bA}_{i-1}\|_{\rm F}^2\nonumber\\
        &+ \mu^2 \delta^2(1-\delta)^2 \mathrm{Tr} \big( \bE \big[\blambda_{i-1} \R_{\bL} \blambda_{i-1}^\bT\big] \big),
        \label{eq:th1_rec}
    \end{align}
    where the following bound holds:
    \begin{align}
        \mathrm{Tr} \big( \bE \blambda_{i-1} \R_{\bL} \blambda_{i-1}^\bT \big) &\leq |\mathcal N| \lambda_{\max}(\R_{\bL}) \lambda_{\max} (\bE \blambda_{i-1}\blambda_{i-1}^\bT).
        \label{eq:th1helper3}
    \end{align}
    Let
    \begin{align}
        \gamma_i = \delta^2\kappa_i|\mathcal N|  \lambda_{\max}(\R_{\bL}).
        \label{eq:gamma_helper}
    \end{align}
    Then, using (\ref{eq:th1_rec}) and (\ref{eq:th1helper3}) we get 
    \begin{align}
        \bE\|\widetilde{\bA}_i\|_{\rm F}^2 \leq \alpha_{i-1}\bE\|\widetilde{\bA}_{i-1}\|_{\rm F}^2 + \mu^2\gamma_i.
        \label{eq:recursion_helper}
    \end{align}
    
    From Appendix~\ref{appendix_lem} it follows that  $\lim_{i\rightarrow\infty}\bE\blambda_i\blambda_i^\bT$ is positive-definite and convergent. Therefore,
    \begin{align}
        \nu \triangleq \lim_{i\rightarrow\infty} \nu_i,\nonumber\\
        \kappa \triangleq \lim_{i\rightarrow\infty}\kappa_i
    \end{align}
    are positive and finite. Thus, according to~(\ref{eq:alpha_helper}) and (\ref{eq:gamma_helper}),
    \begin{align}
        &\alpha \triangleq \lim_{i\rightarrow\infty}\alpha_i = 1 - 2\mu\nu+ O(\mu^2),\nonumber\\
        &\gamma \triangleq \lim_{i\rightarrow\infty}\gamma_i = \delta^2 \kappa |\mathcal N| \lambda_{\max}(\R_{\bL})
    \end{align}
    are both positive, and $\alpha$ is bounded away from one.
    By the definition of a sequence limit, $\forall \varepsilon_1, \varepsilon_2 > 0$, there exist $N(\varepsilon_1, \varepsilon_2) \in \mathbb N$ such that $\forall n \geq N(\varepsilon_1, \varepsilon_2)$:
    \begin{align}
        &|\alpha_n - \alpha| \leq \varepsilon_1,\nonumber\\
        &|\gamma_n - \gamma| \leq \varepsilon_2.
        \label{eq:upper_restrictions}
    \end{align}
    In the following, we will consider $\varepsilon_1$ and $\varepsilon_2$ bounded from above: 
    \begin{align}
        &\varepsilon_1 < \max\{\alpha, 1-\alpha\}\nonumber\\
        &\varepsilon_2 < \gamma
    \end{align}
    so that $(\alpha - \varepsilon_1, \alpha + \varepsilon_1) \subset (0, 1)$ and $\gamma -\varepsilon_2 > 0$.
    We fix $n \geq N(\varepsilon_1, \varepsilon_2)$ and take $i > n$. Then,~(\ref{eq:recursion_helper}) becomes
    \begin{align}
        &\bE\|\widetilde{\bA}_i\|_{\rm F}^2 \leq \alpha_{i-1}\bE\|\widetilde{\bA}_{i-1}\|_{\rm F}^2 + \mu^2\gamma_i \nonumber\\
        \leq&\; (\alpha+\varepsilon_1)^{i-n}\bE\|\widetilde{\bA}_{n}\|_{\rm F}^2 + \mu^2\left(\gamma + \varepsilon_2\right)\sum_{j=0}^{i-n}(\alpha + \varepsilon_1)^j.
    \end{align}
    Since $n$ is fixed and $\alpha + \varepsilon_1 \in (0, 1)$, the steady-state performance can be characterized by
    \begin{align}
        \limsup_{i\rightarrow\infty} \bE\|\widetilde{\bA}_i\|_{\rm F}^2 \leq \frac{\mu^2(\gamma+\varepsilon_2)}{1 - \alpha - \varepsilon_1}.
    \end{align}
    The derived upper bound is independent of $n$, and the analysis applies restrictions on $\varepsilon_1$ and $\varepsilon_2$~(\ref{eq:upper_restrictions}) only from above. Thus, $\varepsilon_1$ and $\varepsilon_2$ can be chosen arbitrary small. Finally, 
    \begin{align}
        \limsup_{i\rightarrow\infty} \bE\|\widetilde{\bA}_i\|_{\rm F}^2 < \frac{\mu^2\gamma}{1 - \alpha}.
    \end{align}
    
\section{Proof of Lemma~\ref{lem:belief}}\label{l3}
    Consider the probability 
    \begin{align}
        p_{k,i} = \mathbb P ( \arg\max_{\theta\in\Theta} \boldsymbol\psi_{k,i} (\theta)\neq\theta^\star )
    \end{align} of choosing the wrong hypothesis for each agent $k\in\mathcal N$ at time instant $i \geq 1$. Without loss of generality, suppose that we construct $\blambda_i$~(\ref{eq:lambda}) with $\theta_0 = \theta^\star$. Then, clearly, 
    \begin{align}
        p_{k,i} = \mathbb P \left(\exists j: [\blambda_i]_{k,j} \leq 0 \right) \leq \sum_{j=1}^{|\Theta|-1} \mathbb P \left( [\blambda_i]_{k,j} \leq 0 \right).
    \end{align}
    As $i\rightarrow 0$, the above inequality is transformed to:
    \begin{align}
        p_{k} \leq \sum_{j}\mathbb P \left([\blambda]_{k,j} \leq 0 \right),
    \end{align}
    where $p_{k} \triangleq \lim_{i\rightarrow\infty} p_{k,i}$, and $\blambda = \lim_{i\rightarrow\infty} \blambda_i$ is given by Lemma~\ref{lem:lambda}. Each element of $\blambda$ has the following form:
    \begin{align}
        [\blambda]_{k,j} = \delta\sum_{t=0}^\infty (1-\delta)^t\sum_{\ell} [A_\star^t]_{\ell,k}[\bL_t]_{\ell,j} = \sum_{\ell \in \mathcal N_k} \boldsymbol s^{(\ell,k)}(\theta_j)
    \end{align}
    with 
    \begin{align}
        &\boldsymbol z_{t}^{(\ell)}(\theta_j) \triangleq \boldsymbol [\bL_t]_{\ell, j}\nonumber\\
        &\alpha_t^{(\ell,k)} \triangleq [A_\star^t]_{\ell, k}\nonumber\\
        &\boldsymbol s^{(\ell,k)}(\theta_j) =  \delta\sum_{t=0}^\infty (1-\delta)^t \alpha_t^{(\ell,k)} \boldsymbol z^{(\ell)}_t(\theta_j).\label{eq:partial_sum}
    \end{align}
    Since $\lim_{i\rightarrow\infty}(A_\star^{\mathsf{T}})^i=\mathds{1}u^{\mathsf{T}}$, where $u$ is the Perron eigenvector~\cite{Sayed_2014}, the sequence $\alpha_t^{(\ell)}$ converges to $0 < \lim_{t\rightarrow\infty}\alpha_t^{(\ell)} \leq 1$. The random variables $\boldsymbol z_{t}^{(\ell)}(\theta_j)$ are i.i.d. w.r.t. time $t$, and their expectations are non-negative with at least one positive element (Assumption~\ref{asm:ident}). To finish the proof, we refer to the original theorem \cite[Theorem 2]{bordignon2020adaptive}) proof, where the behavior of partial sums $\boldsymbol s^{(\ell,k)}(\theta_j)$ of form~(\ref{eq:partial_sum}) is studied. 
    
\section{}\label{apx:influences}
    \noindent Iterating~(\ref{eq:recursion_adaptive}) we have:
    \begin{align}
        \blambda_i = \left(1-\delta\right)^i \left(A_\star^i\right)^\bT\Lambda_0 + \delta \sum_{t=0}^{i-1} \left(1-\delta\right)^{t}(A_\star^{t})^\bT\bL_{i-t}.
    \end{align}
    Each element is represented by:
    \begin{align}
        [\blambda_i]_{k,j} =\;& \left(1-\delta\right)^i \sum_{k'\in\mathcal N}[A_\star^i]_{k',k}[\Lambda_0]_{k',j} \nonumber\\
        &\;+ \delta \sum_{t=0}^{i-1} \left(1-\delta\right)^{t}[A_\star^{t}]_{k',k}[\bL_{i-t}]_{k',j}.
    \end{align}
    For $t<i$, consider the derivative
    \begin{align}
        \frac{\partial [\blambda_i]_{k,j}}{\partial [\bL_{t}]_{\ell, j}} = \delta (1-\delta)^{i-t} \sum_{p_{\ell,k}(v_1,\dots,v_{i-t-1})} a_{\ell,v_1}\dots a_{v_{i-t-1,k}},
    \end{align}
    where the sum is taken over all possible paths of length $it'$ from node $\ell$ to node $k$. To avoid confusion, we clarify the case when $i=t$:
    \begin{align}
        \frac{\partial [\blambda_i]_{k,j}}{\partial [\bL_{i}]_{\ell, j}} = \begin{cases} \delta a_{\ell, \ell},& \textrm{ if } \ell = k,\nonumber\\
        0, & \textrm{ otherwise}.
        \end{cases}
    \end{align}
    Thus, the influence $\eta_d(\ell, k)$ has the following representation:
    \begin{align}
        &\eta_d(\ell, k) = \sum_{j = 1}^{|\Theta|-1} \sum_{r=1}^d \frac{\partial[\blambda_d]_{k,j}}{\partial[\bL_{r}]_{\ell, j}} \nonumber\\
        =&\; (|\Theta|-1) \delta \sum_{r = 0}^{d-1} (1-\delta)^{r} \sum_{v_1,\dots,v_r\in\mathcal N} a_{\ell,v_1}\dots a_{v_{r,k}}.
    \end{align}
    
\section{}\label{app:path}
    We rewrite the optimization problem using the definition of path influence~(\ref{eq:path_influence}):
    \begin{align}
        p_{\ell,k}^\star =&\; \arg\max_{p\in \mathcal P^d_{\ell,k}} \left\{   I\left(p\right) \right\} \nonumber\\
        =&\; \arg\max_{p\in \mathcal P^d_{\ell,k}} \Bigg\{\log\Bigg( (|\Theta|-1)\delta(1-\delta)^r \prod_{(\ell',k')\in p} a_{\ell'k'} \Bigg)\Bigg\}\nonumber\\
        =&\; \arg\min_{p\in \mathcal P^d_{\ell,k}} \Bigg\{ -\log(|\Theta|-1) -\log\delta  - r \log (1-\delta) \nonumber\\
        &\;\;\;\;\;\;\;\;\;\;\;\;\;\;\;\;\;\;\;- \sum_{(\ell',k')\in p} \log a_{\ell'k'} \Bigg\} \nonumber\\
        =&\; \arg\min_{p\in P_{\ell,k}} \Bigg\{ - \sum_{(\ell',k')\in p} \big(\log a_{\ell'k'} + \log\left(1-\delta\right)\big) \Bigg\},
    \end{align}
    where $r$ is the length of a particular path $p$.
    
\bibliographystyle{IEEEtran}
\bibliography{references}

\begin{thebibliography}{10}
\providecommand{\url}[1]{#1}
\csname url@samestyle\endcsname
\providecommand{\newblock}{\relax}
\providecommand{\bibinfo}[2]{#2}
\providecommand{\BIBentrySTDinterwordspacing}{\spaceskip=0pt\relax}
\providecommand{\BIBentryALTinterwordstretchfactor}{4}
\providecommand{\BIBentryALTinterwordspacing}{\spaceskip=\fontdimen2\font plus
\BIBentryALTinterwordstretchfactor\fontdimen3\font minus
  \fontdimen4\font\relax}
\providecommand{\BIBforeignlanguage}[2]{{%
\expandafter\ifx\csname l@#1\endcsname\relax
\typeout{** WARNING: IEEEtran.bst: No hyphenation pattern has been}%
\typeout{** loaded for the language `#1'. Using the pattern for}%
\typeout{** the default language instead.}%
\else
\language=\csname l@#1\endcsname
\fi
#2}}
\providecommand{\BIBdecl}{\relax}
\BIBdecl

\bibitem{shumovskaia}
V.~Shumovskaia, K.~Ntemos, S.~Vlaski, and A.~H. Sayed, ``Online graph learning
  from social interactions,'' in \emph{Proc. Asilomar Conference on Signals,
  Systems, and Computers}, Pacific Grove, CA, 2021, pp. 1263--1267.

\bibitem{jadbabaie2012non}
A.~Jadbabaie, P.~Molavi, A.~Sandroni, and A.~Tahbaz-Salehi, ``Non-bayesian
  social learning,'' \emph{Games and Economic Behavior}, vol.~76, no.~1, pp.
  210--225, 2012.

\bibitem{nedic2017fast}
A.~Nedi{\'c}, A.~Olshevsky, and C.~A. Uribe, ``Fast convergence rates for
  distributed non-bayesian learning,'' \emph{IEEE Transactions on Automatic
  Control}, vol.~62, no.~11, pp. 5538--5553, 2017.

\bibitem{molavi2017foundations}
P.~Molavi, A.~Tahbaz-Salehi, and A.~Jadbabaie, ``Foundations of non-bayesian
  social learning,'' \emph{Columbia Business School Research Paper}, no. 15-95,
  2017.

\bibitem{molavi2018theory}
------, ``A theory of non-bayesian social learning,'' \emph{Econometrica},
  vol.~86, no.~2, pp. 445--490, 2018.

\bibitem{bordignon2020adaptive}
V.~Bordignon, V.~Matta, and A.~H. Sayed, ``Adaptive social learning,''
  \emph{IEEE Transactions on Information Theory}, vol.~67, no.~9, pp.
  6053--6081, 2021.

\bibitem{bordignon2020social}
------, ``Social learning with partial information sharing,'' in \emph{IEEE
  International Conference on Acoustics, Speech and Signal Processing
  (ICASSP)}, Barcelona, Spain, 2020, pp. 5540--5544.

\bibitem{lalitha2018social}
A.~Lalitha, T.~Javidi, and A.~D. Sarwate, ``Social learning and distributed
  hypothesis testing,'' \emph{IEEE Transactions on Information Theory},
  vol.~64, no.~9, pp. 6161--6179, 2018.

\bibitem{zhao2012learning}
X.~Zhao and A.~H. Sayed, ``Learning over social networks via diffusion
  adaptation,'' in \emph{2012 Conference Record of the Forty Sixth Asilomar
  Conference on Signals, Systems and Computers (ASILOMAR)}.\hskip 1em plus
  0.5em minus 0.4em\relax IEEE, 2012, pp. 709--713.

\bibitem{gale2003bayesian}
D.~Gale and S.~Kariv, ``Bayesian learning in social networks,'' \emph{Games and
  Economic Behavior}, vol.~45, no.~2, pp. 329--346, 2003.

\bibitem{acemoglu2011bayesian}
D.~Acemoglu, M.~A. Dahleh, I.~Lobel, and A.~Ozdaglar, ``Bayesian learning in
  social networks,'' \emph{The Review of Economic Studies}, vol.~78, no.~4, pp.
  1201--1236, 2011.

\bibitem{baehrens2010explain}
D.~Baehrens, T.~Schroeter, S.~Harmeling, M.~Kawanabe, K.~Hansen, and K.-R.
  M{\"u}ller, ``How to explain individual classification decisions,'' \emph{The
  Journal of Machine Learning Research}, vol.~11, pp. 1803--1831, 2010.

\bibitem{adadi2018peeking}
A.~Adadi and M.~Berrada, ``Peeking inside the black-box: a survey on
  explainable artificial intelligence ({XAI}),'' \emph{IEEE Access}, vol.~6,
  pp. 52\,138--52\,160, 2018.

\bibitem{gilpin2018explaining}
L.~H. Gilpin, D.~Bau, B.~Z. Yuan, A.~Bajwa, M.~Specter, and L.~Kagal,
  ``Explaining explanations: An overview of interpretability of machine
  learning,'' in \emph{Proc. IEEE Intern. Conference on Data Science and
  Advanced Analytics (DSAA),}.\hskip 1em plus 0.5em minus 0.4em\relax Turin,
  Italy: IEEE, 2018, pp. 80--89.

\bibitem{BARREDOARRIETA202082}
\BIBentryALTinterwordspacing
A.~{Barredo Arrieta}, N.~Díaz-Rodríguez, J.~{Del Ser}, A.~Bennetot, S.~Tabik,
  A.~Barbado, S.~Garcia, S.~Gil-Lopez, D.~Molina, R.~Benjamins, R.~Chatila, and
  F.~Herrera, ``Explainable artificial intelligence (xai): Concepts,
  taxonomies, opportunities and challenges toward responsible ai,''
  \emph{Information Fusion}, vol.~58, pp. 82--115, 2020. [Online]. Available:
  \url{https://www.sciencedirect.com/science/article/pii/S1566253519308103}
\BIBentrySTDinterwordspacing

\bibitem{heuillet2022collective}
A.~Heuillet, F.~Couthouis, and N.~D{\'\i}az-Rodr{\'\i}guez, ``Collective
  explainable {AI}: Explaining cooperative strategies and agent contribution in
  multiagent reinforcement learning with shapley values,'' \emph{IEEE
  Computational Intelligence Magazine}, vol.~17, no.~1, pp. 59--71, 2022.

\bibitem{ohana2021explainable}
J.~J. Ohana, S.~Ohana, E.~Benhamou, D.~Saltiel, and B.~Guez, ``Explainable {AI}
  ({XAI}) models applied to the multi-agent environment of financial markets,''
  in \emph{International Workshop on Explainable, Transparent Autonomous Agents
  and Multi-Agent Systems}.\hskip 1em plus 0.5em minus 0.4em\relax Springer,
  2021, pp. 189--207.

\bibitem{9355456}
J.~Ho and C.-M. Wang, ``Explainable and adaptable augmentation in knowledge
  attention network for multi-agent deep reinforcement learning systems,'' in
  \emph{2020 IEEE Third International Conference on Artificial Intelligence and
  Knowledge Engineering (AIKE)}, 2020, pp. 157--161.

\bibitem{vlaski22}
S.~Vlaski, S.~Kar, A.~H. Sayed, and J.~M.~F. Moura, ``Networked signal and
  information processing,'' \emph{arXiv:2210.13767}, 2022.

\bibitem{wang2019influence}
T.~Wang, J.~Wang, Y.~Wu, and C.~Zhang, ``Influence-based multi-agent
  exploration,'' \emph{arXiv preprint arXiv:1910.05512}, 2019.

\bibitem{Sayed_2014}
\BIBentryALTinterwordspacing
A.~H. Sayed, ``Adaptation, learning, and optimization over networks,''
  \emph{Foundations and Trends® in Machine Learning}, vol.~7, no. 4-5, pp.
  311--801, 2014. [Online]. Available:
  \url{http://dx.doi.org/10.1561/2200000051}
\BIBentrySTDinterwordspacing

\bibitem{Sayed_2023}
------, \emph{Inference and Learning from Data}.\hskip 1em plus 0.5em minus
  0.4em\relax Cambridge University Press, 2023, vols. 1--3.

\bibitem{kalofolias2016learn}
V.~Kalofolias, ``How to learn a graph from smooth signals,'' in
  \emph{Artificial Intelligence and Statistics}.\hskip 1em plus 0.5em minus
  0.4em\relax PMLR, 2016, pp. 920--929.

\bibitem{egilmez2017graph}
H.~E. Egilmez, E.~Pavez, and A.~Ortega, ``Graph learning from data under
  laplacian and structural constraints,'' \emph{IEEE Journal of Selected Topics
  in Signal Processing}, vol.~11, no.~6, pp. 825--841, 2017.

\bibitem{vlaski2018online}
S.~Vlaski, H.~P. Maretić, R.~Nassif, P.~Frossard, and A.~H. Sayed, ``Online
  graph learning from sequential data,'' in \emph{Proc. IEEE Data Science
  Workshop (DSW)}, Lausanne, Switzerland, 2018, pp. 190--194.

\bibitem{dong2019learning}
X.~Dong, D.~Thanou, M.~Rabbat, and P.~Frossard, ``Learning graphs from data: A
  signal representation perspective,'' \emph{IEEE Signal Processing Magazine},
  vol.~36, no.~3, pp. 44--63, 2019.

\bibitem{pasdeloup2017characterization}
B.~Pasdeloup, V.~Gripon, G.~Mercier, D.~Pastor, and M.~G. Rabbat,
  ``Characterization and inference of graph diffusion processes from
  observations of stationary signals,'' \emph{IEEE Transactions on Signal and
  Information Processing over Networks}, vol.~4, no.~3, pp. 481--496, 2017.

\bibitem{thanou2017learning}
D.~Thanou, X.~Dong, D.~Kressner, and P.~Frossard, ``Learning heat diffusion
  graphs,'' \emph{IEEE Transactions on Signal and Information Processing over
  Networks}, vol.~3, no.~3, pp. 484--499, 2017.

\bibitem{chepuri2017learning}
S.~P. Chepuri, S.~Liu, G.~Leus, and A.~O. Hero, ``Learning sparse graphs under
  smoothness prior,'' in \emph{IEEE International Conference on Acoustics,
  Speech and Signal Processing (ICASSP)}, 2017, pp. 6508--6512.

\bibitem{shafipour2017network}
R.~Shafipour, S.~Segarra, A.~G. Marques, and G.~Mateos, ``Network topology
  inference from non-stationary graph signals,'' in \emph{IEEE International
  Conference on Acoustics, Speech and Signal Processing (ICASSP)}, 2017, pp.
  5870--5874.

\bibitem{segarra2016network}
S.~Segarra, A.~G. Marques, G.~Mateos, and A.~Ribeiro, ``Network topology
  identification from spectral templates,'' in \emph{IEEE Statistical Signal
  Processing Workshop (SSP)}, Palma de Mallorca, Spain, 2016, pp. 1--5.

\bibitem{viola2018graph}
I.~Viola, H.~P. Maretic, P.~Frossard, and T.~Ebrahimi, ``A graph learning
  approach for light field image compression,'' in \emph{Applications of
  Digital Image Processing XLI}, vol. 10752.\hskip 1em plus 0.5em minus
  0.4em\relax International Society for Optics and Photonics, 2018, p. 107520E.

\bibitem{sardellitti2016graph}
S.~Sardellitti, S.~Barbarossa, and P.~Di~Lorenzo, ``Graph topology inference
  based on transform learning,'' in \emph{IEEE Global Conference on Signal and
  Information Processing (GlobalSIP)}, Greater Washington, D.C., USA, 2016, pp.
  356--360.

\bibitem{maretic2017graph}
H.~P. Maretic, D.~Thanou, and P.~Frossard, ``Graph learning under sparsity
  priors,'' in \emph{IEEE International Conference on Acoustics, Speech and
  Signal Processing (ICASSP)}, New Orleans, LA, USA, 2017, pp. 6523--6527.

\bibitem{shi2019bayesian}
S.~Shi, G.~Bottegal, and P.~M. Van~den Hof, ``Bayesian topology identification
  of linear dynamic networks,'' in \emph{IEEE European Control Conference
  (ECC)}, Naples, Italy, 2019, pp. 2814--2819.

\bibitem{shahrampour2013reconstruction}
S.~Shahrampour and V.~M. Preciado, ``Reconstruction of directed networks from
  consensus dynamics,'' in \emph{IEEE American Control Conference}, Washington,
  DC, 2013, pp. 1685--1690.

\bibitem{hassan2016topology}
S.~Hassan-Moghaddam, N.~K. Dhingra, and M.~R. Jovanovi{\'c}, ``Topology
  identification of undirected consensus networks via sparse inverse covariance
  estimation,'' in \emph{IEEE Conference on Decision and Control (CDC)}, Las
  Vegas, NV, 2016, pp. 4624--4629.

\bibitem{liu2019dynamic}
C.~Liu, J.~He, S.~Zhu, and C.~Chen, ``Dynamic topology inference via external
  observation for multi-robot formation control,'' in \emph{Pacific Rim
  Conference on Communications, Computers and Signal Processing (PACRIM)},
  Auckland, New Zealand, 2019, pp. 1--6.

\bibitem{ma2008mining}
H.~Ma, H.~Yang, M.~R. Lyu, and I.~King, ``Mining social networks using heat
  diffusion processes for marketing candidates selection,'' in
  \emph{Proceedings of the 17th ACM conference on Information and knowledge
  management}, 2008, pp. 233--242.

\bibitem{friedman2008sparse}
J.~Friedman, T.~Hastie, and R.~Tibshirani, ``Sparse inverse covariance
  estimation with the graphical lasso,'' \emph{Biostatistics}, vol.~9, no.~3,
  pp. 432--441, 2008.

\bibitem{matta2019graph}
V.~Matta, A.~Santos, and A.~H. Sayed, ``Graph learning with partial
  observations: Role of degree concentration,'' in \emph{IEEE International
  Symposium on Information Theory (ISIT)}, Paris, France, 2019, pp. 1312--1316.

\bibitem{sayed2014diffusion}
A.~H. Sayed, ``Diffusion adaptation over networks,'' in \emph{Academic Press
  Library in Signal Processing}, R.~Chellapa and S.~Theodoridis, Eds.\hskip 1em
  plus 0.5em minus 0.4em\relax Academic Press, Elsevier, 2014, vol.~3, pp.
  323--454.

\bibitem{horn2009}
R.~A. Horn and C.~R. Johnson, \emph{Matrix Analysis}.\hskip 1em plus 0.5em
  minus 0.4em\relax Cambridge University Press, NY, 2013.

\bibitem{lalitha2016social}
A.~Lalitha, T.~Javidi, and A.~D. Sarwate, ``Social learning and distributed
  hypothesis testing,'' \emph{IEEE Transactions on Information Theory},
  vol.~64, no.~9, pp. 6161--6179, 2018.

\bibitem{michotte2017perception}
A.~Michotte, \emph{The perception of causality}.\hskip 1em plus 0.5em minus
  0.4em\relax Routledge, 2017, vol.~21.

\bibitem{dijkstra1959note}
E.~W. Dijkstra, ``A note on two problems in connexion with graphs,''
  \emph{Numerische mathematik}, vol.~1, no.~1, pp. 269--271, 1959.

\bibitem{chen2003dijkstra}
J.-C. Chen, ``Dijkstra’s shortest path algorithm,'' \emph{Journal of
  formalized mathematics}, vol.~15, no.~9, pp. 237--247, 2003.

\bibitem{matta2020graphlearning}
V.~Matta, A.~Santos, and A.~H. Sayed, ``Graph learning under partial
  observability,'' \emph{Proceedings of the IEEE}, vol. 108, no.~11, pp.
  2049--2066, 2020.

\end{thebibliography}
\end{document}